\documentclass[11pt]{article}
 \usepackage{titlesec}
\usepackage{hyperref}

\titleclass{\subsubsubsection}{straight}[\subsection]

\usepackage[utf8]{inputenc}
\usepackage{amsmath,amsfonts,amssymb,stackengine,graphicx}

\usepackage[utf8]{inputenc}
\newif\iffigs\figstrue

\usepackage[title]{appendix}
\usepackage{epsfig,latexsym}
\usepackage{hyperref}
\usepackage{amsmath}
\usepackage{verbatim}
\usepackage{color}
\usepackage{mathrsfs}  
\usepackage{slashed}
\usepackage{amssymb}

\DeclareMathAlphabet{\mathpzc}{OT1}{pzc}{m}{it}

 \csname
@addtoreset\endcsname{equation}{section}
  
\def\gz0{\gamma^{0}}

\def\sign{\rm sign}

\def\m{\mu}

\def\beq{\begin{equation}}
\def\eeq{\end{equation}}
\def\bea{\begin{eqnarray}}
\def\eea{\end{eqnarray}}
\def\ba{\begin{array}}
\def\ea{\end{array}}
\def\bec{\begin{center}}
\def\ec{\end{center}}
\def\ba{\begin{align}}
\def\ena{\end{align}}

\def\12{\frac{1}{2}}

\newcounter{subsubsubsection}[subsubsection]
\renewcommand\thesubsubsubsection{\thesubsubsection.\arabic{subsubsubsection}}

\titleformat{\subsubsubsection}
  {\normalfont\normalsize\bfseries}{\thesubsubsubsection}{1em}{}
\titlespacing*{\subsubsubsection}
{0pt}{3.25ex plus 1ex minus .2ex}{1.5ex plus .2ex}

\makeatletter
\renewcommand\paragraph{\@startsection{paragraph}{5}{\z@}%
  {3.25ex \@plus1ex \@minus.2ex}%
  {-1em}%
  {\normalfont\normalsize\bfseries}}
\renewcommand\subparagraph{\@startsection{subparagraph}{6}{\parindent}%
  {3.25ex \@plus1ex \@minus .2ex}%
  {-1em}%
  {\normalfont\normalsize\bfseries}}
\def\toclevel@subsubsubsection{4}
\def\toclevel@paragraph{5}
\def\toclevel@paragraph{6}
\def\l@subsubsubsection{\@dottedtocline{4}{7em}{4em}}
\def\l@paragraph{\@dottedtocline{5}{10em}{5em}}
\def\l@subparagraph{\@dottedtocline{6}{14em}{6em}}
\makeatother

\begin{document}

\thispagestyle{empty}

\begin{flushright}
{\today}
\end{flushright}

\vspace{10pt}

\begin{center}


{\Large\sc A 4D IIB Flux Vacuum and Supersymmetry Breaking}\vskip 12pt
{\large \sc I.~Fermionic Spectrum}


\vspace{25pt}
{\sc J.~Mourad${}^{\; a}$  \ and \ A.~Sagnotti${}^{\; b}$\\[15pt]

${}^a$\sl\small APC, UMR 7164-CNRS, Universit\'e   Paris Cit\'e\\
10 rue Alice Domon et L\'eonie Duquet \\75205 Paris Cedex 13 \ FRANCE
\\ e-mail: {\small \it
mourad@apc.univ-paris7.fr}\vspace{10
pt}

{${}^b$\sl\small
Scuola Normale Superiore and INFN\\
Piazza dei Cavalieri, 7\\ 56126 Pisa \ ITALY \\
e-mail: {\small \it sagnotti@sns.it}}\vspace{10pt}
}

\vspace{40pt} {\sc\large Abstract}\end{center}
\noindent
{We consider the type--IIB supergravity vacua that include an internal $T^5$, depend on a single coordinate $r$ and respect a four--dimensional Poincar\'e symmetry, with the aim of highlighting low--energy spectra with broken supersymmetry and a bounded string coupling.
    These vacua are characterized by the flux $\Phi$ of the self-dual five form in the internal torus,
    the  length $\ell$ of the interval described by the coordinate $r$, a dilaton profile that is inevitably \emph{constant} and a strictly positive dimensionless parameter $h$. As $\ell\rightarrow\infty$ while retaining finite values for $\Phi$ and $\ell \,h^{-\,\frac{5}{4}}$, \emph{half} of the original ten--dimensional supersymmetry is recovered, while finite values of $\ell$ break it completely. In the large--$\ell$ limit one boundary disappears but the other is still present, and is felt as a BPS orientifold by a probe brane.
    In this paper we focus on the fermionic zero modes and show that, although supersymmetry is broken for finite values of $\ell$, they are surprisingly those of four--dimensional $N=4$ supergravity coupled to five vector multiplets. Still, the gravitini can acquire masses via radiative corrections, absorbing four of the massless spin--$\frac{1}{2}$ modes.}

\vskip 12pt

\setcounter{page}{1}

\pagebreak

\newpage
\tableofcontents
\newpage
\baselineskip=20pt
\section{\sc  Introduction and Summary}\label{sec:intro}

Different options for the breaking to minimal $N=1$ supersymmetry in String Theory~\cite{strings} have been scrutinized over the years. The most widely explored ones rest on Calabi--Yau compactifications of the heterotic string~\cite{heterotic}, possibly in the presence of internal fluxes~\cite{grana}, or on their orbifold~\cite{orbifolds} limits. To these one can add type--IIB~\cite{IIB,bergshoeff} orientifolds~\cite{orientifolds}, also leading to $N=1$ supersymmetry, or scenarios that transcend string perturbation theory, based on M-theory~\cite{witten95} and F-theory~\cite{Ftheory}. In retrospect, however, all these pathways proved under control since the residual supersymmetry protects the vacuum. On the other hand, the final breaking to $N=0$, which is needed to grant a proper connection to the Standard Model, can only be addressed within the low--energy supergravity~\cite{supergravity} and entails a subtle back-reaction, which is still not fully under control.

There are, however, different settings that can lead directly to the complete breaking of supersymmetry. Notable among these is the Scherk--Schwarz mechanism~\cite{scherkschwarz}, which naturally connects to no--scale supergravity models~\cite{noscale}. In its different incarnations for closed--string models~\cite{scherkschwarz_closed} and when open strings are also present~\cite{scherkschwarz_open}, it leads to examples where supersymmetry breaking is induced, in an explicit fashion, on entire string spectra, but the vacuum is again severely affected. In addition, the tachyon--free string models of~\cite{so1616,susy95,bsb} provide another, perhaps more fundamental setting, for the complete breaking of supersymmetry. The vacuum suffers once more from severe back-reactions, and tadpole potentials that emerge starting from the (projective) disk order deform the original ten--dimensional Minkowski space. The phenomenon of ``brane supersymmetry breaking''~\cite{bsb} is particularly intriguing, since it embodies a non--linear realization of ten--dimensional supersymmetry~\cite{nonlinearsusy}, a peculiar and surprising feature to be found in String Theory at this level. Still, the Dudas--Mourad vacua~\cite{dm_vacua} yield spontaneous compactifications for these models on string--scale internal spaces, which are driven by the very tadpole potentials, with finite values of Planck mass and gauge couplings in one lower dimension. Strong coupling and curvature singularities are present in some regions, and yet the tools of low--energy supergravity indicate that these vacua are perturbatively stable~\cite{bms}, in contrast to the more symmetrical tadpole--driven AdS vacua of~\cite{adsnonsusy}. Moreover, a mild instability of isotropy present in their cosmological counterparts of~\cite{dm_vacua,climbing} resonates with a possible dynamical origin for a compactified space time. Generalizations of these vacua, also combining fluxes with finite intervals, were recently explored in~\cite{ms21_2}.

Compactifications with fluxes that combine tori and Minkowski spaces with an interval can induce supersymmetry breaking, even in the absence of tadpole potentials. A number of options were recently explored in~\cite{ms21_1,ms21_2}, and in some cases the string coupling can be bounded from above, even in the presence of tadpole potentials. Here we elaborate on what is perhaps the simplest and most interesting option, internal geometries of this type for the type--IIB string~\cite{IIB,bergshoeff} with fluxes of the self--dual Ramond-Ramond five form, which lead to four--dimensional Minkowski space times with broken supersymmetry. A constant dilaton profile is a viable option, thanks to the special features of the five--form field strength, which does not couple to the dilaton field and, as we shall see, the boundary conditions at the ends of the interval~\cite{boundary} make this choice inevitable. 
These models are characterized by the five--form flux $\Phi$ on the internal torus, the length $\ell$ of the interval and a strictly positive dimensionless parameter $h$~\footnote{This class of backgrounds becomes singular as $h\to 0$, and therefore we focus on non--vanishing values for $h$.} that determines the spectrum of massive excitations, and thus the supersymmetry breaking scale, as
\beq{}{}{}{}{}{}{}{}{}{}{}{}{}{}{}{}{}{}{}{}{}
\mu_S \ = \ \frac{1}{\ell\, h^\frac{1}{4} } \ . \label{susyb_scale}
\eeq
Supersymmetry can be recovered in the limit $\mu_S \to 0$, while retaining finite values for $\Phi$ and for the combination $\ell\,h^{-\frac{5}{4}}$, and therefore as both $\ell$ and $h$ tend to infinity. For finite values of $\ell$, the resulting low--energy effective field theory is four dimensional and has a finite Planck mass
\beq
m_{Pl(4)}^2 \ \sim \  m_{Pl(10)}^8\ \frac{\ell^2\,\Phi}{\sqrt{h}} \ .
\eeq
Note that here the volume of the internal torus is a derived quantity, in contrast to the standard Kaluza--Klein scenario, where it would be a modulus. 

Our main task, here and in the companion papers~\cite{ms22_2}, is a detailed analysis of the mode spectra arising from this class of compactifications. We have split the analysis into three parts, for clear reasons of relative brevity, but not only. Here we discuss the salient properties of the backgrounds and analyze the resulting spectra of Fermi modes, relying to a large extent on links with Schr\"odinger--like systems. On the other hand, in the first of~\cite{ms22_2} we shall discuss the spectra of Bose modes and their indications for the perturbative stability of these vacua. However, we shall leave aside the sector of singlet scalar modes with nonzero toroidal momenta, which presents some peculiar and unexpected features, and will be the subject of the third paper. Bose modes require a detailed analysis of mixings of Kaluza--Klein excitations, and drawing some lessons on the stability of the resulting spectra will also require different techniques.

In the following, we summarize the contents of this paper in detail.

In Section~\ref{sec:background}, following~\cite{ms21_1}, we describe the vacua of type--IIB supergravity with an internal five-torus and four--dimensional Poincar\'e symmetry. A convenient choice for the coordinate $r$
    on which this class of backgrounds depends leads to exact solutions of the low--energy equations, and we show that a constant axion-dilaton profile is inevitable.
    As we have anticipated, these vacua are characterised by the parameters $\Phi$, $\ell$ and $h$, and for $\ell$ finite the resulting space-times have everywhere a finite string coupling and include a finite internal interval. Still, curvature singularities are present at its ends, so that, while string loop corrections can be held under control, $\alpha'$ string corrections are expected to be important, at least within their neighborhoods.
    However, within a wide region of parameter space, where $\ell\,M_s>>1$ and $\Phi\, M_s^4>>1$, with $M_s \sim \frac{1}{\sqrt{\alpha'}}$ the string scale, even these $\alpha'$ corrections become negligible within a sizable fraction of the internal interval.
    In this Section we also show that finite values of $\ell$ break all supersymmetries, while in the $\ell\rightarrow \infty$ limit half of the original ten--dimensional supersymmetries are recovered, albeit in a five--dimensional warped spacetime. In the supersymmetric limit the interval becomes a half-line, so that one boundary is still present, and we show how a probe brane reveals that the endpoint behaves as an $\overline{O}_-$ orientifold. This type of behavior is approached asymptotically near the origin in all these vacua.
    
   In Section~\ref{sec:Fermi modes} we study the massless Fermi modes of the background. The boundaries raise the question of what conditions one should impose there on the different fields. Our analysis relies on our previous work~\cite{boundary}, which is further adapted to this setting in Appendix~\ref{app:fermi_interval}: we thus demand that the charges associated with infinitesimal translations in spacetime and along the internal torus, together with those associated to spacetime Lorentz symmetries, be unable to flow across the boundaries. However, there is some freedom beyond the options that we identified in~\cite{boundary}, where we focused implicitly on orientifold models~\cite{orientifolds}. The presence of pairs of identical Fermi fields in the massless spectrum of the type--IIB string allows in fact to mix them, so that the boundary conditions
   \beq{}{}{}{}{}{}{}
   \left(\Lambda \ \pm \ 1 \right)\psi \ = \ 0 \
   \eeq
  can be solved, compatibly with all residual symmetries in the geometries of interest, by
   \beq{}{}{}{}{}{}{}
   \Lambda \ = \ i\, \gamma^0\,\gamma^1\,\gamma^2\,\gamma^3\, \sigma_2 \ .
   \eeq
   This choice combines the four--dimensional chirality matrix with the $\sigma_2$ Pauli matrix, mixing pairs of type--IIB Fermi fields compatibly with their Majorana--Weyl nature. { We focus largely on identical boundary conditions at the two ends, which allow the presence of massless Fermi modes, but we also discuss other options.}

   Mode normalisation is a key issue: acceptable (Fermi or Bose) modes are to be normalizable, in order to acquire a four--dimensional interpretation. This poses no constraints for the internal torus, but it does for the interval, whose parametrization involves an infinite range in $r$. Therefore, it is crucial to identify the proper normalization integrals in order to select the modes of interest. In simple cases, these conditions can be deduced rather directly from the low--energy effective field theory, but when mixings are present matters become more subtle. In these cases the reduction to Schr\"odinger--like systems will prove our key tool in this respect. Our main result is that{, if identical boundary conditions are enforced at the ends of the internal interval,} the tree--level fermionic zero modes comprise a quartet of massless Majorana four--dimensional gravitini and six quartets of Majorana spin--$\frac{1}{2}$ fermions. These build altogether the fermionic content of $N=4$ supergravity coupled to five vector multiplets, despite the breaking of supersymmetry.
  
  Section~\ref{sec:conclusions} summarizes our results and elaborates on some perspectives for future work, and includes a discussion on how the gravitini can acquire masses by radiative corrections. Appendix~\ref{app:fermiEinstein} collects a number of technical details on the equations of motion of the different types of fermionic modes present in these backgrounds.    Appendix~\ref{app:fermi_interval} summarizes and completes the discussion of the boundary problem for fermions in an interval presented in~\cite{boundary}.
Finally, Appendix~\ref{app:sturmliouville} elaborates on Schr\"odinger--like formulations for Bose and Fermi fields.

\section{\sc The Background} \label{sec:background}

In this section we derive the explicit form of the background profiles of interest and analyze some of the resulting physical properties. Our basic requirement is to end up with a four--dimensional Minkowski background, and for simplicity we allow a non--trivial dependence on only one internal coordinate $r$, thus focusing on a class of type--IIB background metrics of the form
\beq
ds^2 \ = \ e^{2 A(r)} dx^2 \ + \ e^{2B(r)} dr^2 \ + \ e^{2 C(r)} dy^2 \ . \label{metric}
\eeq
Here the $x$ coordinates refer to a four--dimensional Minkowski space time, while the $y$ coordinates refer to a compact five--dimensional internal space that, for simplicity, we take to be a torus characterized by a single radius $R$. The non--trivial features of this class of metrics are encoded in the three functions $A(r)$, $B(r)$ and $C(r)$. Notice, however, that $B(r)$ can be changed by reparametrizations, and in the following we shall make the convenient ``harmonic'' gauge choice
\beq
B(r) \ = \ 4 \, A(r) \ + \ 5\,C(r) \ , \label{harmonic}
\eeq
which will lead to handy exact solutions. The class of vacua that we shall explore is sustained by an $r$--dependent self--dual five form of type IIB, whose profile is fully determined by the isometries and the condition of self--duality, and reads~\footnote{In the notation of~\cite{ms21_1,ms21_2} $H=\frac{H_5}{2\sqrt{2}}$.} 
\beq
{\cal H}_5^{(0)} \ = \ {H} \Big( e^{4A+B-5C} \, dx^0 \wedge ...\wedge dx^3\wedge dr\ + \ dy^1 \wedge ... \wedge dy^5 \Big) \ , \label{4d_inter_flux_H}
\eeq
where $H$ is a constant, which we shall often assume to be positive in the following. In addition, one can allow for $r$--dependent profiles for the dilaton--axion of type IIB. To this end, it is convenient to work in the terms of the complex combination
\beq{}{}{}{}{}{}{}{}{}{}{}{}{}{}{}{}{}{}
\tau \ = \ a \ + \ i\,e^{-\phi} \ ,
\eeq
which transforms under SL(2,R) according to
\beq{}{}{}{}{}{}{}{}{}{}{}{}{}{}{}{}{}{}
\tau' \ = \ \frac{\alpha \,\tau \ + \ \beta}{\gamma\,\tau \ + \ \delta}  \ ,
\eeq
with $\alpha$, $\beta$, $\gamma$ and $\delta$ four real parameters subject to the constraint
\beq{}{}{}{}{}{}{}{}{}{}{}{}{}{}{}{}{}{}
\alpha\,\delta \ - \ \beta\,\gamma \ = \ 1 \ .
\eeq
The equations for the axion--dilaton pair follow from the action
\beq{}{}{}{}{}{}{}{}{}{}{}{}{}{}{}{}{}{}
{\cal S} \ = \ - \ \frac{1}{2}\,\int d^{10}\,x\  \sqrt{-g}\ \frac{\partial_M \overline{\tau}\,\partial^M \tau}{\left[{\rm Im}\ \tau\right]^2} , \label{dilaxion_action}
\eeq
and in backgrounds of the type~\eqref{metric}, in the harmonic gauge~\eqref{harmonic} and for scalar profiles only depending on $r$, they reduce to
\beq
\left(a'\,e^{2\phi}\right)'\ = \ 0 \ , \qquad \phi'' \, - \,  \left(a'\right)^2\, e^{2\phi} \ = \ 0 \ . \label{dilaxion_eqs}
\eeq
On the other hand, in backgrounds of the form~\eqref{metric} the Einstein equations
\beq
R_{MN}  =  \frac{1}{24}\ {{\cal H}_{5}^{(0)}}_{M PQRS}\, {{\cal H}_{5}^{(0)}}_{N P'Q'R'S'} g^{PP'} g^{QQ'}g^{RR'}g^{SS'}+ \frac{1}{4}\frac{\partial_M\,\tau\,\partial_N\,\bar{\tau} + \partial_N\,\tau\,\partial_M\,\bar{\tau}}{\left[{\rm Im}\ \tau\right]^2}  \label{eqs_sdual}
\eeq
reduce to
\bea
&& A'' \ = \  {H^2}\ e^{8 A} \ , \nonumber \\
&& C'' \ = \  - \ {H^2}\ e^{8 A}\ , \nonumber\\
&& 3 \left(A'\right)^2 \ + \ 10\, A'\, C' \ + \ 5 \left(C'\right)^2 \ = \ - \ \frac{H^2}{2} \ e^{8 A} \ + \ \frac{1}{8}\,\frac{\tau'\,\bar{\tau}'}{\left[{\rm Im}\ \tau\right]^2} \  .  \label{background_eqs2}
\eea

\subsection{\sc Derivation of the Background Solution}

In this section we provide some details on the derivation of the background solution, and on the inevitable emergence of a constant dilaton profile in this setting.

\subsubsection{\sc The $SL(2,R)/U(1)$ Scalar Sector}

To begin with, the first of eqs.~\eqref{dilaxion_eqs} is solved letting
\beq
a'\ = \ c\, e^{\,-\,2\phi} \ ,
\eeq
where $c$ is a constant, and the equation for $\phi$ then takes the familiar form~\cite{ms21_1,ms21_2}
\beq
\phi'' \ = \ c^2 \, e^{-\,2\,\phi} \ . \label{eq_dilaton}
\eeq
It is convenient to distinguish two cases:
\begin{enumerate}
    \item if $c=0$, the solution of eq.~\eqref{eq_dilaton} is
    \beq{}{}
    \phi \ = \ \phi_1\,r \ + \ \phi_2 \ , \label{lin_dil}
    \eeq
    for arbitrary values of the two constants $\phi_1$ and $\phi_2$, and the axion profile $a$ is constant;
    \item if $c\neq 0$, using some results discussed, for instance, in Appendix B of \cite{ms21_1}, the solution of eq~\eqref{eq_dilaton} reads
    \beq{}{}
    e^\phi \ = \ \frac{c}{\widetilde{\phi}_1} \ \cosh\left(\widetilde{\phi}_1\,r \ + \ \widetilde{\phi}_2\right) \ , \label{lin_dil2}
    \eeq
\end{enumerate}
where $\widetilde{\phi}_1$ and $\widetilde{\phi}_2$ are two constants and now $c \,\widetilde{\phi}_1 > 0$. Notice that, in the limit where $\widetilde{\phi}_2$ is very large, eq.~\eqref{lin_dil2} reduces to eq.~\eqref{lin_dil} with $\phi_1=\widetilde{\phi}_1 \neq 0$ and $e^{\phi_2} = \frac{c}{2\,\widetilde{\phi}_1}\,e^{\widetilde{\phi}_2}$. Therefore, finite values of $\phi_2$ can be recovered in the limit of vanishing $c$, and this suggests to present eq.~\eqref{lin_dil2} in the equivalent form
\beq
 e^\phi \ = \ 2\,e^{\phi_2-\widetilde{\phi}_2} \ \cosh\left({\phi}_1\,r \ + \ \widetilde{\phi}_2\right) \ ,
\eeq
which can encompass all cases and rests on the three constants $\phi_1$, $\phi_2$ and $\widetilde{\phi}_2$. The string coupling takes its lowest value, on the whole real axis, where the argument of the hyperbolic function vanishes. If $\phi_1=0$, which is relevant for the supersymmetric case, the axion and dilaton profiles are both constant.

Eq.~\eqref{lin_dil2} then determines
\beq{}{}{}{}{}{}{}{}{}{}{}{}{}{}{}{}{}{}
\tau \ = \ \frac{e^{ \widetilde{\phi}_2-\phi_2}}{2\,\cosh\left(\phi_1\,r\,+\,\widetilde{\phi}_2\right)}\left[ \sinh\left(\phi_1\,r\,+\,\widetilde{\phi}_2\right) \ + \ i \right] \ + \ a_0 \ , \label{dilaxion_sol}
\eeq
where $a_0$ is a constant. Notice that for $r \to \pm\,\infty$
the step--wise axion profile approaches constant values,
\beq{}{}{}{}{}{}{}{}{}{}{}{}{}{}{}{}{}{}
\frac{1}{2}\, \sign\left(\phi_1\,r\right)\,e^{\widetilde{\phi}_2-\phi_2} \ + \ a_0\ ,
\eeq
while its derivative decays exponentially, and the dilaton approaches the linear behavior
\beq{}{}{}{}{}{}{}{}{}{}{}{}{}{}{}{}{}{}
\phi \ \sim \phi_1\,r \ + \ \phi_2 \ .
\eeq
Note also that the scalar contribution to the Einstein equations in the form~\eqref{eqs_sdual}, which is only present in the $rr$ component, is simply $\frac{1}{2}\,\phi_1^2$, and is insensitive to the values of the other three constants, which are mapped into one another by $SL(2,R)$ transformations.

Summarizing, the axion--dilaton profile can be determined exactly in the harmonic gauge, as in~\eqref{dilaxion_sol}, up to a few constants and independently of the actual values of $A(r)$, $B(r)$ and $C(r)$, up to the boundary conditions that will emerge shortly.

\subsubsection{\sc The Metric Profiles}

For the class of backgrounds of interest, in the harmonic gauge~\eqref{harmonic} the field equations reduce to eqs.~\eqref{background_eqs2}. The last of them, the ``Hamiltonian constraint", reduces the independent integration constants.

Adding the first two equations one can see that
\beq
C' \ =\  - \ A' \ - \ \alpha \ , \label{CA}
\eeq
with $\alpha$ a constant. Making use of this result in the Hamiltonian constraint then gives
\beq
\left(A'\right)^2 \ - \ \frac{H^2}{4}\,e^{8A} \ = \ \frac{5}{2}\,\alpha^2 \ - \ \frac{1}{16}\, (\phi_1)^2 \ .
\eeq

Eqs.~\eqref{background_eqs2} are then all satisfied, and letting
\beq
Y \ = \ e^{-4A} \label{Y_A}
\eeq
one is left with
\beq
\left( Y' \right)^2 \ - \ E \, Y^2 \ = \ 4\,H^2 \ ,
\eeq
where
\beq
E \ = \ 40\,\alpha^2 \ -\, \phi_1^2 \ .
\eeq
There are three families of solutions, depending of the value of $E$:
\begin{itemize}
    \item  if $E>0$, letting
    \beq{}{}{}{}{}{}{}{}{}{}{}{}{}
    \frac{1}{\rho} \ = \ \sqrt{E} \ .
    \eeq
the solutions read
\beq
e^{-4 A} \ = \ Y(r) \ = \ {2 \left|H\right|\,\rho}\,\sinh\left(\frac{r}{\rho}\right) \ ,
\eeq
with $0< r< \infty$. In this case for $\alpha$ and $\phi_1$ there are two branches of solutions, which can be parametrized, via a real parameter $\zeta$, as
\beq{}{}{}{}{}{}{}{}{}{}{}{}{}
\alpha \ = \ \pm \ \frac{1}{2\rho\sqrt{10}}\, \cosh \zeta \ , \qquad \phi_1 \ = \ \frac{1}{\rho}\,\sinh \zeta \ .
\eeq
Consequently
\bea
B &=& - \ A \ \mp \ \frac{\sqrt{10}}{4\rho}\ r\, \cosh \zeta \ + \ 5\,\beta \ , \nonumber \\
C &=& - \ A \ \mp \ \frac{\sqrt{10}}{20\rho}\ r\,\cosh \zeta \ + \ \beta \ ,
\eea
where $\beta$ is another integration constant, and the background takes the form
\bea{}{}{}{}{}{}{}{}{}{}{}{}{}
ds^2 &=& \left[ \frac{1}{2\left|H\right|\rho\,\sinh\left(\frac{r}{\rho}\right)}\right]^\frac{1}{2} dx^2 \nonumber \\ &+& \left[{2\left|H\right|\,\rho}\,\sinh\left(\frac{r}{\rho}\right)\right]^\frac{1}{2} \left(e^{ \mp \ \frac{\sqrt{10}}{2\rho}\, r\, \cosh \zeta \ + \ 10\,\beta} \, dr^2 \ + \ e^{\mp \ \frac{\sqrt{10}}{10\rho}\, r\,\cosh \zeta \,+\,2\,\beta} \, d\vec{y}^2\right) \ , \nonumber \\
\tau &=& a+ i\,e^{\,-\,\phi} \ = \ \frac{e^{\widetilde{\phi}_2-\phi_2}}{2\,\cosh\left(\frac{\sinh\zeta\,r}{\rho}\,+\,\widetilde{\phi}_2\right)}\left[ \sinh\left(\frac{\sinh\zeta\,r}{\rho}\,+\,\widetilde{\phi}_2\right) \ + \ i \right] \,+\, a_0\ , \nonumber \\
{{\cal H}_5^{(0)}} &=&
H \left\{ \frac{dx^0 \wedge ...\wedge dx^3\wedge dr}{\left[2\,{|H|}\,\rho\,\sinh\left(\frac{r}{\rho}\right)\right]^2} \ + \ dy^1 \wedge ... \wedge dy^5\right\} \ ; \label{back_epos}
\eea
\item  if $E=0$, there are again two branches of solutions, which read
\beq
Y(r) \ = \ 2 \left|H\,r\right| \ ,
\eeq
with $0< r< \infty$,
and
\beq{}{}{}{}{}{}{}{}{}{}{}{}{}
\phi_1 \ = \ \pm\ 2\sqrt{10}\ \alpha \ ,
\eeq
and the background takes the form
\bea{}{}{}{}{}{}{}{}{}{}{}{}{}
ds^2 &=& \frac{dx^2}{\Big(2\left|H\right| r\Big)^\frac{1}{2}} \ + \ \Big(2\left|H\right| r\Big)^\frac{1}{2} \left(e^{ \mp \ \frac{\sqrt{10}}{2}\, \phi_1\, r \ + \ 10\,\beta} \, dr^2 \ + \ e^{\mp \ \frac{\sqrt{10}}{10}\, \phi_1\, r\,+\,2\,\beta} \, d\vec{y}^2\right) \ , \nonumber \\
\tau &=& a \ + \ i\,e^{-\phi}\ = \ \frac{e^{\widetilde{\phi}_2-\phi_2}}{2\,\cosh\left(\phi_1\,r\,+\,\widetilde{\phi}_2\right)}\Big[ \sinh\left(\phi_1\,r\,+\,\widetilde{\phi}_2\right) \ + \ i \Big] \,+\, a_0 \ , \nonumber \\
{{\cal H}_5^{(0)}} &=&
H \left\{ \frac{dx^0 \wedge ...\wedge dx^3\wedge dr}{\Big(2 \left|H\right| \,r\Big)^2} \ + \ dy^1 \wedge ... \wedge dy^5\right\} \ . \label{back_ezero}
\eea
These results can be obtained as limits of the preceding ones;
\item if $E<0$, letting
    \beq{}{}{}{}{}{}{}{}{}{}{}{}{}
    \frac{1}{\rho} \ = \ \sqrt{\left|E\right|} \ ,
    \eeq
\beq
Y(r) \ = \ 2\left|H\right|\,\rho\,\sin\left(\frac{r}{\rho}\right) \ ,
\eeq
with $0< r< \pi\,\rho $, and now there are two branches of solutions described by
\beq{}{}{}{}{}{}{}{}{}{}{}{}{}
\alpha \ = \ \frac{1}{2\rho\sqrt{10}}\, \sinh \zeta \ , \qquad \phi_1 \ = \ \pm \ \frac{1}{\rho}\, \cosh \zeta \ ,
\eeq
while the background takes the form
\bea{}{}{}{}{}{}{}{}{}{}{}{}{}
ds^2 &=& \left[ \frac{1}{2\left|H\right|\rho\,\sin\left(\frac{r}{\rho}\right)}\right]^\frac{1}{2} dx^2 \nonumber \\ &+& \left[2\left|H\right|\,\rho\,\sin\left(\frac{r}{\rho}\right)\right]^\frac{1}{2} \left(e^{\,-\,\frac{\sqrt{10}}{2\rho}\, r\,\sinh \zeta \ + \ 10\,\beta} \, dr^2 \ + \ e^{ \,-\, \frac{\sqrt{10}}{10\rho}\,r\, \sinh \zeta\,+\,2\,\beta} \, d\vec{y}^2\right) \ , \nonumber \\
\tau &=& a+ i\,e^{\,-\,\phi} \ = \ \frac{e^{\widetilde{\phi}_2-\phi_2}}{2\,\cosh\left(\pm \frac{\cosh\zeta\,r}{\rho}\,+\,\widetilde{\phi}_2\right)}\left[ \sinh\left(\pm \frac{\cosh\zeta\,r}{\rho}\,+\,\widetilde{\phi}_2\right) \ + \ i \right] \,+\, a_0 \ , \nonumber \\
{{\cal H}_5^{(0)}} &=&
H \left\{ \frac{dx^0 \wedge ...\wedge dx^3\wedge dr}{\left[2\left|H\right|\rho\,\sin\left(\frac{r}{\rho}\right)\right]^2} \ + \ dy^1 \wedge ... \wedge dy^5\right\} \ . \label{back_eneg}
\eea
\end{itemize}
Notice that the metric background depends only on $\phi_1$, while it is independent of $\phi_2$ and $\widetilde{\phi}_2$.  In all the preceding expressions the $y$'s are periodic coordinates, with period $2\pi R$.

\subsubsection{\sc Boundary Conditions}

An important feature of all these backgrounds is the presence of a singularity at $r=0$ and, in one case, of an additional singularity at a finite value of $r$. Moreover, for $E \geq 0$ even $r=+\infty$ can lie at a finite distance from the origin. We shall see shortly that these are non--trivial singularities, so that spacetime includes an interval with one or two boundaries.

The issue is now whether or not the preceding solutions satisfy proper boundary conditions. For gravity, the equations that we have obtained follow from the Einstein--Hilbert action supplemented by the Gibbons--Hawking term~\cite{ght}, while for the form one can verify that there are no $T^{\mu r}$ components of the energy--momentum tensor, which would enter the conditions in~\cite{boundary}. On the other hand, the scalar equations~\eqref{dilaxion_eqs} require a more detailed discussion, since they follow from the action~\eqref{dilaxion_action}, if the boundary conditions
\beq{}{}{}{}{}{}{}{}{}{}{}{}{}{}{}{}{}{}{}{}
\sqrt{-g}\,\left. T^{r}{}_{\mu}\right|_{\partial\,{\cal M}} \ \equiv \ \left. \frac{\overline{\tau}\,\tau'}{\left[{\rm Im}\ \tau\right]^2}\right|_{\partial\,{\cal M}} \ = \ 0 \
\eeq
hold. However, for the general dilaton--axion profile in eq.~\eqref{dilaxion_sol}
\beq{}{}{}{}{}{}{}{}{}{}{}{}{}{}{}{}{}{}{}{}
\frac{\overline{\tau}\,\tau'}{\left[{\rm Im}\ \tau\right]^2} \ = \ \phi_1\left\{ 2\,a_0\,e^{\phi_2-\widetilde{\phi}_2} \ - \ i\left[\cosh\left(\phi_1\,r\,+\,\widetilde{\phi_2}\right) + 2 a_0\sinh\left(\phi_1\,r\,+\,\widetilde{\phi_2}\right)\right]\right\} \ ,
\eeq
and setting $\phi_1=0$ is the only way of satisfying the boundary condition at $r=0$. Consequently, the axion--dilaton profile is bound to be constant. This removes the solutions for $E<0$ and poses no further restrictions on the others, so that one is left with two classes of solutions:
\begin{enumerate}
    \item for $E>0$
    \bea{}{}{}{}{}{}{}{}{}{}{}{}{}
ds^2 &=& \left[ \frac{1}{2\left|H\right|\rho\,\sinh\left(\frac{r}{\rho}\right)}\right]^\frac{1}{2} dx^2 \nonumber \\ &+& \left[2\left|H\right|\,\rho\,\sinh\left(\frac{r}{\rho}\right)\right]^\frac{1}{2} \left(e^{ \mp \ \frac{\sqrt{10}}{2\rho}\, r \ + \ 10\,\beta} \, dr^2 \ + \ e^{\mp \ \frac{\sqrt{10}}{10\rho}\, r \,+\,2\,\beta} \, d\vec{y}^2\right) \ , \nonumber \\
\tau &=& a_0+ i\,e^{\,-\,\phi_0} \ , \nonumber \\
{{\cal H}_5^{(0)}} &=&
H\left\{ \frac{dx^0 \wedge ...\wedge dx^3\wedge dr}{\left[2 \left|H\right|\,\rho\,\sinh\left(\frac{r}{\rho}\right)\right]^2} \ + \ dy^1 \wedge ... \wedge dy^5\right\} \ ; \label{back_epos_fin}
\eea
\item for $E=0$
\bea{}{}{}{}{}{}{}{}{}{}{}{}{}
ds^2 &=& \frac{dx^2}{\Big(2\left|H\right| r\Big)^\frac{1}{2}} \ + \ \Big(2\left|H\right|r \Big)^\frac{1}{2} \left(e^{10\,\beta} \, dr^2 \ + \ e^{2\,\beta} \, d\vec{y}^2\right) \ , \nonumber \\
\tau &=& a_0 \ + \ i\,e^{-\phi_0} \ , \nonumber \\
{{\cal H}_5^{(0)}} &=&
H \left\{ \frac{dx^0 \wedge ...\wedge dx^3\wedge dr}{\Big(2 \left|H\right| \,r\Big)^2} \ + \ dy^1 \wedge ... \wedge dy^5\right\} \ . \label{back_ezero_fin}
\eea
\end{enumerate}
Actually, the solutions of the second type are a limiting case of those of the first type, and are recovered as $\rho \to \infty$ or as $r \to 0$.

\subsection{\sc Physical Properties of the Background} \label{sec:background2}

The two residual families of backgrounds of eqs.~\eqref{back_epos_fin} and \eqref{back_ezero_fin} depend apparently on $\rho$, $H$, $\beta$ and $\phi_2$. Moreover, they also depend on the radii of the internal $T^5$, which we take to be all identical and equal to $R$ for simplicity, and in the former case also on a discrete choice of branch.

\subsubsection{\sc Canonical Forms of the Solutions}

It is now instructive to perform some redefinitions in eqs.~\eqref{back_epos}. For later convenience, we thus let
\beq{}{}{}{}{}{}
h \ = \ 2\,H\,\rho \ ,  \label{hH5}
\eeq
use the new variables
\beq{}{}{}{}{}{}
\tilde{r} \,=\, \frac{r}{\rho} \ , \qquad
\tilde{y}^i \,=\, \frac{y^i}{2\pi R}   \label{xry}
\eeq
and introduce the length scale
\beq{}{}{}{}{}{}
\ell \,=\, \rho \ h^\frac{1}{4}\,e^{-5\beta} \label{ell h}
\eeq
and the five-form flux in the internal torus, which according to eq.~\eqref{4d_inter_flux_H} is
\beq{}{}{}{}{}{}
\Phi \ = \ H \, \left(2\pi R\right)^5  \ = \ \frac{\left(2\pi\,h^\frac{1}{4}\,R\right)^5}{{2}\,\ell}\ e^{-5\beta} \ .
\eeq

In terms of these new variables, the solutions with $E>0$ become (we drop ``tilde's'' for brevity, while also warning the reader that we are using the same symbol $r$ for a coordinate that has been rescaled, and is now dimensionless)
\bea
ds^2 &=& \frac{\eta_{\mu\nu}\,dx^\mu\,dx^\nu}{\left[h\,\sinh\left(r\right)\right]^\frac{1}{2}} \,+\, \ell^2 \,\left[\sinh\left(r\right)\right]^\frac{1}{2}\, e^{ \mp \ \frac{\sqrt{10}}{2}\, r} \, dr^2 \nonumber \\ &+& \left({2}\,\Phi\,\ell\right)^\frac{2}{5}\left[\sinh\left({r}\right)\right]^{\,\frac{1}{2}} \  e^{\mp \ \frac{\sqrt{10}}{10}\, r}   \,d\vec{y}^{\,2}\ , \nonumber \\
\tau &=& a_0+ i\,e^{\,-\,\phi_0} \ , \nonumber \\
{{\cal H}_5^{(0)}} &=&
\frac{1}{2\,h}\, \frac{dx^0 \wedge ...\wedge dr}{\left[\sinh \left({r}\right)\right]^2} \ + \ \Phi\, dy^1 \wedge ... \wedge dy^5  \ . \label{4d_PhiEpos}
\eea
Note that with these new coordinates the background no longer depends on $r$, while the ``harmonic'' gauge condition~\eqref{harmonic} becomes
\beq{}{}{}{}{}{}{}{}{}{}{}
e^B \ = \ \frac{h}{{2}\,\Phi} \ e^{4A+5C} \ .
\eeq

On the other hand, for $E=0$ one can perform the redefinitions
\beq{}{}{}{}{}{}
\tilde{r} \,=\, 2\,H\,r \ , \qquad
\tilde{y}^i \,=\, \frac{y^i}{2\pi R} \ ,  \label{xry_zero}
\eeq
which turn the background into
\bea{}{}{}{}{}{}{}{}{}{}{}{}{}
ds^2 &=& \frac{dx^2}{r^\frac{1}{2}} \ + \ r^\frac{1}{2} \left[e^{10\,\beta} \, \frac{dr^2}{4\,H^2} \ + \ e^{2\,\beta} \, \left(2 \pi R \right)^2\,d\vec{y}^2\right] \ , \nonumber \\
\tau &=& a_0 \ + \ i\,e^{-\phi_0} \ , \nonumber \\
{{\cal H}_5^{(0)}} &=&
\frac{1}{2} \, \frac{dx^0 \wedge ...\wedge dx^3\wedge dr}{r^2} \ + \ \Phi \,dy^1 \wedge ... \wedge dy^5 \ , \label{back_ezero_scale}
\eea
where we drop again ``tilde's'' for brevity. The two final redefinitions
\beq{}{}{}{}{}{}{}{}{}{}{}{}{}
e^{10\beta}\,\left(\frac{1}{2\,H\,\ell}\right)^2 \ = \ 1 \ , \qquad x \ \rightarrow \ h^{-\,\frac{1}{4}}\,x
\eeq
turn the background into a form along the lines of the other cases,
\bea{}{}{}{}{}{}{}{}{}{}{}{}{}
ds^2 &=& \frac{dx^2}{\sqrt{h\,r}} \ + \ r^\frac{1}{2} \left[\ell^2\,dr^2 \ + \ \left({2}\,\Phi\,\ell\right)^\frac{2}{5}\,d\vec{y}^2\right] \ , \nonumber \\
\tau &=& a_0 \ + \ i\,e^{-\phi_0} \ , \nonumber \\
{{\cal H}_5^{(0)}} &=&
\frac{1}{2\,h} \, \frac{dx^0 \wedge ...\wedge dx^3\wedge dr}{r^2} \ + \ \Phi \,dy^1 \wedge ... \wedge dy^5 \ . \label{back_ezero_scale_2}
\eea
Note that, referring to $\Phi$, $\ell$ and $h$, $\beta$ and the scale $R$ have completely disappeared from the problem, in all cases. Note also that the flux $\Phi$ should be quantized~\cite{witten_flux} according to
\beq
q_3\,\Phi \ = \ n \ , \label{Dirac}
\eeq
where $N$ is an integer and $q_3$ is the D3-brane charge~\cite{strings}
\beq{}{}{}{}{}{}
q_3 \ = \ \sqrt{\pi}\,m_{Pl(10)}^4 \ . \label{q3}
\eeq

\subsubsection{\sc Internal Length and Effective Planck Mass}

The length of the $r$-interval is finite for a subset of the solutions, the upper branch with $E>0$ in eqs.~\eqref{4d_PhiEpos}, for which
\beq
L \ = \ \int_0^\infty e^B \, dr \ = \ \ell  \, \int_0^\infty dx \ e^{\,-\,\frac{5\,x}{2\sqrt{10}} } \,\left(\sinh x\right)^\frac{1}{4} \, \simeq \  1.43\, \ell \ . \label{ell1}
\eeq
On the other hand, the length is infinite in the two other cases in eqs.~\eqref{4d_PhiEpos} and \eqref{back_ezero_scale_2}.

The corresponding behaviors of the Planck mass are determined by
\beq
m_{Pl(4)}^2 \,=\, m_{Pl(10)}^8\,\int\, dr \, d^{\,5} y \, \sqrt{-g} \ e^{-2A}\,=\, \frac{2\,m_{Pl(10)}^8\,\Phi}{h} \, \int\, dr \ e^{2(B-A)} \ ,
\eeq
and the $r$ integral is finite only for the upper branch of solutions with $E>0$. In our solutions
\beq{}{}{}{}{}{}{}
m_{Pl(10)}^8 \ = \ \frac{1}{\left(\alpha'\right)^4\,g_s^2} \ = \ \frac{1}{\left(\alpha'\right)^4}\ e^{-2\phi_0}\ , \label{mpl84}
\eeq
with $\alpha'$ is the Regge slope, since the string coupling $g_s$ is $e^{\phi_0}$.

For the upper branch of the $E>0$ solutions in eqs.~\eqref{4d_PhiEpos} the effective Planck mass is finite, and is given by
\bea
m_{Pl(4)}^2 &=&  \frac{4 \,m_{Pl(10)}^8\,\ell^2\,\Phi}{3\,\sqrt{h}} \ . \label{planck4}
\eea

Since our analysis rests on the effective field theory, the results can be reliable in String Theory only if the Kaluza--Klein excitations in the $r$-interval and in the internal torus are much lighter than string modes. These conditions translate into the inequalities
\beq{}{}{}{}{}
\frac{\ell}{\sqrt{\alpha'}}\ \gg \ 1  \ , \qquad \frac{\left({2}\,\Phi\,\ell\right)^\frac{1}{5}}{\sqrt{\alpha'}} \ \gg \ 1 \ , \label{bounds}
\eeq
which also grant that one can ignore winding modes on the internal torus. Once the first holds, in general the second inequality does not impose stringent conditions on the flux $\Phi$, and thus on the quantum number $n$. Taking into account eqs.~\eqref{mpl84} and \eqref{planck4}, one can see that these conditions are not incompatible with small values of $g_s$.

Summarizing, we have found three types of solutions, which are all encompassed by two equivalent forms. The first presentation of the background depends on the four parameters $H$, $\rho$, $R$ and $\phi_0$, and the coordinates $y^i$ of the internal $T^5$ that have range $2\pi R$, and also on $a_0$, which however can be removed by an $SL(2,R)$ transformation. It reads
    \bea{}{}{}{}{}{}{}{}{}{}{}{}{}
ds^2 &=& \frac{dx^2}{\left[2\left|H\right|\rho\,\sinh\left(\frac{r}{\rho}\right)\right]^\frac{1}{2}} \nonumber \\ &+& \left[2 \left|H\right|\,\rho\,\sinh\left(\frac{r}{\rho}\right)\right]^\frac{1}{2} \left(e^{ \,-\,\epsilon \, \frac{\sqrt{10}}{2\rho}\, r} \, dr^2 \ + \ e^{\,-\,\epsilon \, \frac{\sqrt{10}}{10\rho}\, r} \, d\vec{y}^2\right) \ , \nonumber \\
\tau &=& a_0+ i\,e^{\,-\,\phi_0} \ , \nonumber \\
{{\cal H}_5^{(0)}} &=&
H\left\{ \frac{dx^0 \wedge ...\wedge dx^3\wedge dr}{\left[2\left|H\right|\,\rho\,\sinh\left(\frac{r}{\rho}\right)\right]^2} \ + \ dy^1 \wedge ... \wedge dy^5\right\} \ , \label{back_epos_fin2}
\eea
where $r>0$, and the values $\epsilon=\pm 1$ distinguish the two branches. The last family of solutions is recovered in the limit $r \to 0$.

The second presentation of the background depends on $h$, $\Phi$ and $\ell$, and the coordinates $y^i$ of the internal $T^5$ have range $1$, and reads
\bea
ds^2 &=& \frac{\eta_{\mu\nu}\,dx^\mu\,dx^\nu}{\left[h\,\sinh\left(r\right)\right]^\frac{1}{2}} \,+\, \ell^2 \,\left[\sinh\left(r\right)\right]^\frac{1}{2}\, e^{ \,-\,\epsilon \ \frac{\sqrt{10}}{2}\, r} \, dr^2 \nonumber \\ &+& \left({2}\,\Phi\,\ell\right)^\frac{2}{5}\left[\sinh\left({r}\right)\right]^{\,\frac{1}{2}} \  e^{\,-\,\epsilon \ \frac{\sqrt{10}}{10}\, r}   \,d\vec{y}^{\,2}\ , \nonumber \\
\tau &=& a_0+ i\,e^{\,-\,\phi_0} \ , \nonumber \\
{{\cal H}_5^{(0)}} &=&
\frac{1}{{2\,h}}\, \frac{dx^0 \wedge ...\wedge dr}{\left[\sinh \left({r}\right)\right]^2} \ + \ \Phi\, dy^1 \wedge ... \wedge dy^5  \ , \label{4d_PhiEpos2}
\eea
and the solutions of the second type are again recovered in the limit $r \to 0$.

Among the three classes of backgrounds, the solutions with $\epsilon=1$ and $E>0$ stand out as physically more interesting since, as we have seen, they lead to a compactification to four dimensions with a finite Planck mass and a bounded string coupling. In the following, we shall largely concentrate on them.

\subsubsection{\sc Limiting Behavior and Singularities}

The limiting behavior as $r\to 0$ of these backgrounds is identical for all types of solutions, which approach the $E=0$ case of eqs.~\eqref{back_ezero_scale}. In this limit $\ell$ plays no role, and could be removed completely rescaling the $r$ coordinates and the spacetime variables $x^\mu$. In terms of the proper length, defined as
\beq{}{}{}{}{}{}{}
\xi \ = \ \frac{4}{5}\,r^\frac{5}{4} \ ,
\eeq
the limiting form of the background becomes
\bea
ds^2 &=&\frac{\eta_{\mu\nu}\,dx^\mu\,dx^\nu}{\sqrt{h}\left(\frac{5}{4}\,\xi\right)^\frac{2}{5}} \,+\, \ell^2 \,d\xi^2 \,+\, \left(\frac{5}{2}\ \xi\,\Phi\,\ell\right)^\frac{2}{5} \,d\vec{y}^{\,2}\ \ , \nonumber \\
{{\cal H}_5^{(0)}} &=&
\frac{1}{{2\,h}}\, \frac{dx^0 \wedge ...\wedge d\xi}{\left(\frac{5}{4}\,\xi\right)^\frac{9}{5}} \ + \ \Phi\, dy^1 \wedge ... \wedge dy^5  \ .
\label{4d_inter_flux_H_app_ell2_xi}
\eea
Hence, as $\xi \to 0$ the scale factor in spacetime becomes unbounded, the scale factor in the internal torus tends to zero and the components of the five-form along spacetime and $\xi$ become unbounded.

At the opposite end of the interval, the behavior depends on the type of solution.

\begin{itemize}
    \item For the upper branch with $E>0$ in eqs.~\eqref{back_epos_fin2}, or equivalently in eqs.~\eqref{4d_PhiEpos2}, the length is finite as $r \to \infty$. Letting
\beq{}{}{}{}{}{}{}
a \ = \ \frac{5}{2\sqrt{10}}\,-\,\frac{1}{4} \ \simeq \  0.54 \ , \qquad
b \ = \  \frac{1}{4} \,-\,\frac{1}{2\sqrt{10}} \ \simeq \ 0.09 \ , \label{abzetaupper}
\eeq
one can work in terms of the proper length, whose form is well approximated by
\beq{}{}{}{}{}{}{}
\xi \ = \ \xi_\infty \ - \ \frac{1}{2^\frac{1}{4}\,a}\ e^{\,-\,a\,r}
\eeq
for large values of $r$, with a limiting value $\xi_\infty$ corresponding to the length $L$ in eq.~\eqref{ell1}. Consequently, the limiting behavior of the background is captured by
\bea
ds^2 &=& \sqrt{\frac{2}{h}}\,\left[2^\frac{1}{4}\,a\left(\xi_\infty\ - \ \xi\right) \right]^\frac{1}{2a} \eta_{\mu\nu}\,dx^\mu\,dx^\nu \,+\, {\ell^2}\ d\xi^2 \nonumber \\ &+& \frac{1}{\sqrt{2}} \left({2}\,\Phi\,\ell\right)^\frac{2}{5}\left[2^\frac{1}{4}\,a\left(\xi_\infty\ - \ \xi\right) \right]^{\,-\,\frac{2b}{a}} d\vec{y}^{\,2}\ \ , \nonumber \\
{{\cal H}_5^{(0)}} &=&
\frac{4}{{2}^\frac{3}{4}\,h}\,\left[2^\frac{1}{4}\,a\left(\xi_\infty\ - \ \xi\right) \right]^{\frac{2-a}{a}}\,{dx^0 \wedge ...\wedge d\xi} \ + \ \Phi\, dy^1 \wedge ... \wedge dy^5   \ . \label{4d_inter_flux_H_app_ell_inf}
\eea
As $\xi$ approaches $\xi_\infty$, the scale factor in spacetime tends to zero while the scale factor on the torus becomes unbounded, and the components of the five-form along spacetime and $\xi$ tend to zero.

\item For the lower branch with $E>0$ in eqs.~\eqref{back_epos_fin2}, or equivalently in eqs.~\eqref{4d_PhiEpos2}, the length of the interval is infinite, and letting
\beq{}{}{}{}{}{}{}
a \ = \ \frac{5}{2\sqrt{10}}\,+ \,\frac{1}{4} \ \simeq \ +  1.04 \ , \qquad
b \ = \  \frac{1}{4} \,+\,\frac{1}{2\sqrt{10}} \ \simeq \ 0.41 \ , \label{abzetalower}
\eeq
one can define the proper length, whose form is well approximated by
\beq{}{}{}{}{}{}{}
\xi \ = \ \frac{1}{2^\frac{1}{4}\,a}\ e^{\,a\,r} \ \geq \ 0
\eeq
for large values of $r$, and the limiting behavior of the background for large values of $\xi$ is captured by
\bea
ds^2 &=& \sqrt{\frac{2}{h}}\left[2^\frac{1}{4}\,a\,\xi \right]^{\,-\,\frac{1}{2\,a}}\,\eta_{\mu\nu}\,dx^\mu\,dx^\nu \,+\, {\ell^2}\ d\xi^2 \,+\, \frac{1}{\sqrt{2}} \left({2}\,\Phi\,\ell\right)^\frac{2}{5}\left[2^\frac{1}{4}\,a\,\xi \right]^{\,\frac{2b}{a}} d\vec{y}^{\,2}\ \ , \nonumber \\
{{\cal H}_5^{(0)}} &=&
\frac{4}{{2}^\frac{3}{4}\,h}\,\left[2^\frac{1}{4}\,a\,\xi \right]^{\,-\,\frac{2+a}{a}}\,{dx^0 \wedge ...\wedge d\xi} \ + \ \Phi\, dy^1 \wedge ... \wedge dy^5   \ . \label{4d_inter_flux_H_app_ell_inf2}
\eea
As $\xi$ approaches $+\infty$, as in the previous case the scale factor in spacetime tends to zero while the scale factor on the torus becomes unbounded, and the components of the five-form along spacetime tend to zero.

\item For $E=0$, the limiting behavior as $r\to +\infty$ is also captured by eqs.~\eqref{4d_inter_flux_H_app_ell2_xi} as $\xi \to \infty$. Therefore, as in the previous cases the scale factor in spacetime tends to zero, the scale factor in the internal torus becomes unbounded and the spacetime components of the five--form field strength tend to zero.
\end{itemize}

Note that $r=0$ is a true singularity, since
\beq
{R^{(0)}}_{MN} \,{R^{(0)}}^{MN} \ \sim \ \frac{1}{\ell^4\,r^5} \
\eeq
for all these solutions. As a result, sizable $\alpha'$-corrections are expected in String Theory within that region, while the present classical treatment ought to be reliable for
\beq
r \ > \ \left[\frac{\sqrt{\alpha'}}{\ell}\right]^\frac{4}{5} \ . \label{left_bound}
\eeq

On the other hand, for $E > 0$, as $r \to \infty$
\beq
R_{MNPQ}\,R^{MNPQ} \ \sim \ \frac{1}{\ell^4}\, e^{\,{r}\left(\frac{ 10\,\epsilon\,-\,\sqrt{10}}{\sqrt{10}}\right)} \ , \label{squared_curv}
\eeq
so that for the upper branch of solutions in eqs.~\eqref{4d_PhiEpos}, with $\epsilon=1$, which grant a finite length of the interval and thus a four--dimensional interpretation, one expects small $\alpha'$-corrections to the low--energy effective field theory for
\beq
r \ < \ \log\left(\frac{\ell}{\sqrt{\alpha'}}\right) \ . \label{cond_zeta}
\eeq
This value should be larger than the bound~\eqref{left_bound} in order for the current treatment to have some intermediate domain of validity, which is guaranteed provided
\beq
\ell \ \gg \ \sqrt{\alpha'} \ .
\eeq
In the other cases the interval has an infinite length, so that the second singularity is not relevant.

In order to build phenomenologically interesting scenarios one should contemplate the addition of branes, but we shall refrain from doing it here, contenting ourselves with a detailed analysis of the modes supported by the first type of background and of their indications for its stability, which will the subject of~\cite{ms22_2}.

\subsection{\sc A Probe Brane in the $r$--Interval}\label{app: probe brane}

The effective Lagrangian for a probe $D3$--brane spanning the four--dimensional Minkowski space, with fixed internal coordinates and an $r$ coordinate that evolves in time, is determined by the induced metric and the coupling to the gauge field corresponding to the ${\cal H}_5^{(0)}$ field strength
\beq
{\cal H}_5^{(0)} \ = \ dx^0 \wedge \ldots \wedge dx^3 \wedge b'(r) \, dr  + \star \ ,
\eeq
where $\star$ denotes the Hodge dual and $b$ is a function of $r$ only.
For a background of the form~\eqref{metric} and \eqref{4d_inter_flux_H}, in the gauge~\eqref{harmonic} and in the Einstein frame, the action takes the form
\beq
\frac{{\cal S}}{V_3} \ = \ - T_3 \int \ dt \, e^{4A(r(t))} \, \sqrt{1 \ - \ e^{2(B-A)(r(t))}\,\dot{r}(t)^2} \ + \ q_3\, \int b[r(t)] \ dt \ ,
\eeq
where $T_3$, $q_3$ and $V_3$ are the brane tension, charge and volume. For the solutions with $E>0$ in eqs.~\eqref{back_epos_fin2}, or equivalently in eqs.~\eqref{4d_PhiEpos2}
\beq
b'(r) \ =\ \frac{1}{4\,H}\,\frac{1}{\left[\rho\,\sinh\left(\frac{r}{\rho}\right)\right]^2}\ , \label{eqbprime}
\eeq
so that
\beq
b(r) \ = \ - \ \frac{1}{{4\,\rho\,H}} \left[\coth\left(\frac{r}{\rho}\right) \ - \ 1\right] \ .
\eeq
The corresponding results for the solutions with $E=0$ can be obtained from these in the limit $\rho \to \infty$.

The energy conservation condition for the probe is then
\beq
\frac{T_3\ e^{4A(r(t))}}{\sqrt{1 \ - \ e^{2(3A+5C)(r(t))}\,\dot{r}(t)^2}} \ - \ q_3\, b \ = \ E \ . \label{en_cons}
\eeq
Close to $r=0$ the limiting behavior of the background, as we have seen, is universal, and in the non--relativistic limit the preceding equation becomes
\beq{}{}{}{}{}{}
\frac{T_3}{2}\, \dot{r}^2 \ + \ \frac{1}{2\,|H|\,r}\left[T_3\ + \ \frac{q_3}{2}\, \sign(H)\right] \ = E \ , \label{en_cons_0}
\eeq
from which one can identify the potential
\beq{}{}{}{}{}{}
V \ \sim \  \frac{1}{r}\left[T_3 \ + \ \frac{q_3}{{2}}\,\sign\left(H\right)\right] \ ,
\eeq
up to a positive overall factor.
This potential describes a gravitational repulsion sized by the $T_3$ term and an ``electric'' interaction that is repulsive for $q_3 \,H >0$ and attractive for $q_3\,H <0$~\footnote{Note that here we are referring to the Einstein equations in the form~\eqref{eqs_sdual}, and thus in the conventions of~\cite{bergshoeff}. A standard normalization would thus obtain rescaling the tensor field by a factor $\frac{1}{\sqrt{2}}$, and consequently in the present notation non--relativistic interactions between two objects with tensions $T_1$ and $T_2$ and electric charges $Q_1$ and $Q_2$ are proportional to the combination
\beq{}{}{}{}{}{}{}{}{}{}{}
T_1\,T_2 \ - \ \frac{1}{2}\,Q_1\,Q_2 \ .
\eeq
Comparing with our result leads to eq.~\eqref{BPS}.}. As a result, one can see that the origin behaves as an orientifold that, in our conventions, has a \emph{negative} tension $T$ and a charge $Q$ that has the same sign as $H$, such that
\beq{}{}{}{}{}{}
Q \ = \ - \ T\, \sign(H) \ . \label{BPS}
\eeq

For the upper branch with $E>0$, near the right end of the finite interval the energy conservation condition becomes
\beq{}{}{}{}{}{}
\frac{T_3}{2}\,e^{\,-\,\frac{5\, r}{\rho\,\sqrt{10}}} \, \dot{r}^2\,+\,\frac{T_3}{\rho\,|H|} \,e^{\,-\,\frac{r}{\rho}} \,+\,\frac{q_3}{2\,\rho\,H}\,e^{\,-\,\frac{2\,r}{\rho}} \ \simeq \ E \ .
\eeq
In order to recover a non--relativistic kinetic term as in eq.~\eqref{en_cons_0},
one can perform the change of variable
\beq
\alpha\,\rho\left(1 \ - \ e^{\,-\,\frac{r}{\alpha\,\rho}}\right) \ = \ u \ ,
\eeq
with
\beq{}{}{}{}{}
\alpha \ = \ \frac{2\sqrt{10}}{5} \ \simeq \ {1.26}  \ ,
\eeq
which inverts to
\beq{}{}{}{}{}{}{}
e^{\,-\,\frac{r}{\rho}} \ = \ \left(1 \ - \ \frac{u}{\alpha\,\rho}\right)^\alpha \ .
\eeq
and leads to
\beq{}{}{}{}{}{}
\frac{T_3}{2}\, \dot{u}^2 \, + \, \frac{T_3}{\rho\,|H|} \,\left(1 \ - \ \frac{u}{\alpha\,\rho}\right)^\alpha \,+\,\frac{q_3}{2\,\rho\,H}\left(1 \ - \ \frac{u}{\alpha\,\rho}\right)^{2\alpha} \ \simeq \  E \ . \label{en_cons_inf}
\eeq
One can thus see that the gravitational force attracts the brane toward the right end, on account of the second term above, while the electric force attracts it there for $q_3\,H>0$ and repels it for $q_3\,H<0$, on account of the third term. However, both forces tend to zero as $u$ approaches $\alpha\,\rho$, but are not proportional. One may rightfully wonder about the fate of the electric tensor flux, which seems to wane across the finite interval. In fact, there is no contradiction with the conservation of electric flux, since the solution is precisely the counterpart of a uniform electric field in our metric background, and satisfies
\beq
b' e^{5C-4A-B} \ = \ \Phi \ ,
\eeq
as can be deduced taking the dual of the constant internal components along the torus.

One can gain some qualitative insights on the overall brane motion noting that the energy $E$ is bounded from below by the static potential
\beq
V(r) \ = \ T_3e^{4A}-q_3b\ =\  \frac{1}{2\,|H|\,\rho} \left[ \frac{T_3}{\sinh\left(\frac{r}{\rho}\right)} \ + \ \frac{q_3\,\sign(H)}{2} \left(\coth\left(\frac{r}{\rho}\right) \,-\,1\right)\right] \ , \label{pot_rho}
\eeq
and the brane  has turning points where $E=V(r)$. Notice that the static potential $V$ contains two contributions, which are both singular at $r=0$ and tend to zero as $r\to \infty$. As we have seen, the first contribution, proportional to $T_3$, looks like a gravitational interaction but \emph{repels} the brane from the origin, while the second, proportional to $q_3$, \emph{attracts} it to the origin if $q_3\,\sign(H) <0$ and \emph{repels} it if $q_3\,\sign(H)>0$. Hence, the origin behaves as a BPS orientifold with negative tension and positive or negative charge, depending on the sign of $H$, consistently with the fact that half of the original supersymmetry is preserved there, as we shall see in the next section.

\subsection{\sc Supersymmetric Vacua} \label{sec:killingspinors}

In this section we prove what we already mentioned, namely that the $E=0$ background with constant dilaton profile preserves half of the original 32 supercharges of type IIB. As we shall see, together with flat space this is the only option, within the class of metrics in eq.~\eqref{metric}, where some supersymmetry is left. To this end, one can look systematically for Killing spinors in IIB backgrounds, within the class of metrics
\beq
ds^2 \ = \ e^{2 A(r)} \,dx^2 \ + \ e^{2B(r)}\, dr^2 \ + \ e^{2 C(r)}\, dy^2 \ ,
\eeq
with a generic $r$--dependent dilaton profile and the self--dual tensor field strength
\beq
{\cal H}_5 \ = \ H \left\{ e^{4A+B-5C} \, dx^0 \wedge ...\wedge dx^3 \wedge dr\ + \ dy^1 \wedge ... \wedge dy^5 \right\} \ ,
\eeq
which already appeared in eq.~\eqref{4d_inter_flux_H}. The supersymmetry transformations of the ten--dimensional IIB theory in the presence of non--trivial dilaton and five--form backgrounds can be recast in the convenient form of eqs.~\eqref{susy_final}. Since~\footnote{Had one allowed for an axon profile, it would be also eliminated by this condition.}
\beq{}{}{}{}{}{}{}{}{}{}{}{}{}
\delta\,\lambda \ = \  {\partial}\!\!\!/ \phi\,\epsilon \ ,
\eeq
the constant dilaton profiles selected by the boundary conditions imply the supersymmetry invariance of $\lambda$, and moreover there is no essential distinction between Einstein and string frames in supersymmetric vacua of this type.

Combining eqs.~\eqref{susy_final}, \eqref{Hslash} and \eqref{omegaab}, the remaining Killing--spinor equations reduce to
\bea
\delta\,\psi_r &=& \partial_r \, \epsilon \, + \, \frac{H}{4} \ e^{B-5C}\ \gamma^{0\ldots 3}  \,i\,\sigma_2\, \epsilon \ = \ 0 \ ,
\nonumber \\
\delta\,\psi_\mu &=& \partial_\mu \, \epsilon \,+\, \frac{1}{2}\ \gamma_\mu \gamma_r\ e^{A-B}\, A'\,\epsilon \, + \, \frac{H}{4} \ e^{A\,-\,5\,C}\  \gamma^{0\ldots 3} \gamma_r\,\gamma_\mu \,i\,\sigma_2\,\epsilon  \ = \ 0  \ ,
\nonumber \\
\delta\,\psi_i &=& \partial_i \, \epsilon \,+\, \frac{1}{2}\ \gamma_i \gamma_r\ e^{C-B}\, C'\,\epsilon \, + \, \frac{H}{4} \ e^{\,-\,4\,C}\  \gamma^{0\ldots 3} \gamma_r\,\gamma_i \,i\,\sigma_2\, \epsilon  \ = \ 0  \ ,
\label{killing_spinors3}
\eea
after taking into account the self--dual nature of the tensor field strength and the spinor chirality projections. All $\gamma$ matrices, here and in the following, have flat indices, and
$\epsilon$ is a doublet of ten-dimensional chiral spinors. One can decompose $\epsilon$ into eigenstates $\epsilon_\pm$ of the Hermitian matrix
\beq
\Lambda=\gamma^{0\ldots 3}\,i\,\sigma_2
\eeq
corresponding to its eigenvalues $\pm 1$, and it is also convenient to define
\beq
J'(r) \ = \ \frac{H}{4} \ e^{B-5C} \ , \label{eqJ}
\eeq
so that eqs.~\eqref{killing_spinors3} read
\bea
&&\partial_r\,\epsilon_\pm \ \pm \ J'(r) \, \epsilon_\pm \ = \ 0 \ , \nonumber \\
&&\partial_\mu\, \epsilon_\pm  \ + \ \frac{1}{2} \,\gamma_\mu \gamma_r\ e^{A-B}\left( A' \ \mp \ 2\,J'\right) \epsilon_\mp \ = \ 0 \ , \nonumber \\
&&\partial_i\, \epsilon_\pm  \ + \ \frac{1}{2} \,\gamma_i \gamma_r\ e^{C-B}\left( C' \ \mp \ 2\,J'\right) \epsilon_\pm \ = \ 0 \ . \label{killing_J}
\eea
The first of these equations is solved by
\beq
\epsilon_\pm \ = \ e^{\,\mp\,J(r)}\, \epsilon_{0\,\pm}(x,y) \ ,
\eeq
where $\epsilon_{0\,\pm}$ are arbitrary functions of the space--time coordinates $x$ and the toroidal coordinates $y$, but are independent of $r$.
The remaining equations now reduce to
\bea
&&\partial_\mu\, \epsilon_{0\,\pm}(x,y)  \ + \ \frac{1}{2} \,\gamma_\mu \gamma_r\ e^{A-B\pm 2J}\left( A' \ \mp \ 2\,J'\right) \epsilon_{0\,\mp}(x,y) \ = \ 0 \ , \nonumber \\
&&\partial_i\, \epsilon_{0,\pm}(x,y)  \ + \ \frac{1}{2} \,\gamma_i \gamma_r\ e^{C-B}\left( C' \ \mp \ 2\,J'\right) \epsilon_{0\,\pm}(x,y) \ = \ 0 \ . \label{killeqs}
\eea
For consistency, the $x$--derivatives of the first and the $y$--derivatives of the second imply the conditions
\beq
\left( A'\right)^2  \ - \ 4\,\left(J'\right)^2 \ = \ 0 \ , \qquad \left( C'  \ \mp \ 2\,J'\right)^2\epsilon_{0\,\pm} \ = \ 0 \ ,
\eeq
which are solved by
\beq
A'\ = \ 2\,\varepsilon_A\, J' \ , \qquad  C'\ = \ 2\,\varepsilon_C\, J' \ , \qquad \epsilon_{0\,-\epsilon_C} = 0 \ ,
\eeq
where $\varepsilon_A$ and $\varepsilon_C$ are signs. Moreover, the very form of eqs.~\eqref{killing_J} constrains the two signs $\varepsilon_A$ and $\varepsilon_C$ to be opposite, so that the solutions are finally
\beq
A'\ = \ 2\,\sigma\, J' \ , \qquad  C'\ = \ -\, 2\,\sigma\, J' \ , \qquad \epsilon_{0\,\sigma} = 0 \ , \label{eqsAC}
\eeq
where $\sigma=\pm1$. Eqs.~\eqref{killeqs} then imply that the leftover $\epsilon_0$ is a constant spinor.

Combining these results with the definition~\eqref{eqJ} now leads to the differential equation
\beq
A'\ = \ \frac{\sigma\,H}{2} \ e^{B-5C} \ ,  \label{aprimesusy}
\eeq
whose solution in the Harmonic gauge~\eqref{harmonic} reads
\beq
e^{-4\,A} \ = \ - \ 2\,\sigma\,H\,r \ ,
\eeq
up to a shift of $r$. One can work conveniently in the region $r>0$ taking
\beq
\sigma \ = \ - \ \sign(H) \ ,
\eeq
and the solution of eqs.~\eqref{eqsAC} finally reads
\beq
e^{2\,A} \ = \ e^{\,-\,2\left(C-c_s\right)} \ = \ \left[\frac{1}{2\,|H|\,r}\right]^\frac{1}{2}   \ ,
\eeq
where $c_s$ is a constant that can be scaled out of all the following expressions.
The end results for the metric and the form field strength are thus
\bea
ds^2 &=&  \frac{dx^2}{\left(2\,|H|\,r\right)^\frac{1}{2}}  \, dx^2 \, + \,  \left(2\,|H|\,r\right)^\frac{1}{2} \left(dr^2 \,+\,   dy^2\right) \ , \nonumber \\
{\cal H}_{5} &=& H \left[ \frac{dx^0 \wedge \ldots \wedge dx^3 \wedge dr}{\left(2\,|H|\,r\right)^\frac{1}{2}}  \,+\, dy^1 \wedge \ldots \wedge dy^5 \right]\ . \label{killing_data}
\eea
These are precisely the $E=0$ background of eqs.~\eqref{back_ezero} for $\phi_1=0$, up the irrelevant constant $\beta$, which can be scaled out. Moreover, these results capture the limiting behavior of the solutions in eqs.~\eqref{back_epos_fin2}, or equivalently in eqs.~\eqref{4d_PhiEpos2}, as $r \to 0$. These limiting behaviors approach a supersymmetric background, since, as we have just seen, they are compatible with the existence of the Killing spinor
\beq
\epsilon \ = \  \frac{1}{\left(2\,|H|\,r\right)^\frac{1}{8}}\, \epsilon_{0} \ , \label{killingeps}
\eeq
and thus preserve 16 of the original 32 supersymmetries of ten--dimensional flat space. Here $\epsilon_0$ is a constant spinor subject to the condition
\beq
\Lambda\,\epsilon_0 \ = \ \gamma^{0\ldots 3}\,i\,\sigma_2\, \epsilon_0 \ = \ \sign\left(H\right) \epsilon_0 \ . \label{Lambdasusy}
\eeq
Note that, within the backgrounds of eqs.~\eqref{back_epos_fin2}, the supersymmetric case is recovered in the $\rho \to \infty$ limit. This is consistent with the scale of supersymmetry breaking that we anticipated in the Introduction, which takes the form
\beq{}{}{}{}{}{}{}{}{}{}{}{}{}{}{}{}{}{}{}{}{}
\mu_S \ \sim \ \frac{1}{\rho^\frac{3}{2} \sqrt{H}} 
\eeq
when expressed in terms of $\rho$.

We can now proceed to analyze the fermionic modes present in the first branch of $E>0$ backgrounds in eqs.~\eqref{back_epos_fin2}, or equivalently in eqs.~\eqref{4d_PhiEpos2}, which are characterized by finite values for the length scale $\ell$ of the internal and for the four--dimensional Planck mass, and by a constant profile for the string coupling. From now on, for definiteness, we shall assume that $H>0$.

\section{\sc Four--dimensional Fermi Modes}\label{sec:Fermi modes}

The original type--IIB theory contains a pair of left--handed Majorana--Weyl gravitini $\psi_M$ and a pair of right--handed  Majorana--Weyl dilatini $\lambda$ in ten dimensions, {which will be treated as $SU(2)$ doublets, as in~\cite{bergshoeff}, where the fermionic action and the supersymmetry transformations are presented in the string frame. The counterpart of these results in the Einstein frame, after some convenient field redefinitions, is described in Appendix~\ref{app:fermiEinstein}.

In this section we analyze the nature of the four--dimensional fermionic modes, paying attention to the massless ones, { whenever they are present}. Since the internal manifold has boundaries, our analysis will rely on~\cite{boundary} and on its refinement in Appendix~\ref{app:fermi_interval}. As we show there, the resulting boundary conditions at the ends of the internal interval have in general the form
\beq
\Lambda\,\psi_M \ = \ \pm \, \psi_M \ , \qquad  \Lambda\,\lambda \ = \ \pm \, \lambda \ ,
\eeq
where the Hermitian matrix, which in our case is
\beq
\Lambda \ = \ \gamma^{0123}\,i\,\sigma_2 \ ,  \label{lambdamat}
\eeq
and satisfies the conditions
\beq{}{}{}{}{}{}{}{}{}{}{}{}{}
\left\{\gamma_0\,\gamma_r\,,\,\Lambda\right\} \ = \ 0 \ , \qquad \left[\gamma_{\mu\nu}\,,\,\Lambda\right] \ = \ 0 \,,\, \qquad \left[\gamma_{11}\,,\,\Lambda\right] \ = \ 0 \ , \qquad C^{-1}\,\Lambda^T\,C = -\,\gamma_0\,\Lambda\,\gamma_0 \ .
\eeq
This matrix already emerged in our discussion of the supersymmetric limit, and in particular in eq.~\eqref{Lambdasusy}.

The linearized supergravity equations of motion for the Fermi fields in the backgrounds described in the previous section are determined by the results in Appendix~\ref{app:fermiEinstein}, and read
\bea{}{}{}{}{}{}{}{}{}{}{}{}{}{}{}{}{}{}
&& {\Gamma^{MNP}\,D_N\,\psi_P \ + \ \frac{1}{8}\,\Gamma^{[M}\,{\cal H}\!\!\!\!/ \ \Gamma^{N]}\,i\,\sigma_2\,\psi_N}  \ = \ 0 \ , \label{grav_eq_final_tx}   \nonumber \\
&& \Gamma^M\,D_M\,\lambda  \ + \ \frac{1}{4}\,{\cal H}\!\!\!\!/ \ i\,\sigma_2\,\lambda \ = \ 0 \ . \label{spinor_eqs1_final_text1}
\eea
{ The derivation in Appendix~\ref{app:fermiEinstein}, which combines the fields $\psi_M$ and $\Gamma_M\,\lambda$ of~\cite{bergshoeff}, suggests for $\psi_\mu$ a $\Lambda$ eigenvalue  opposite to those of $\lambda$, $\psi_r$ and $\psi_i$, so that the original variables of~\cite{bergshoeff} and our new fields obey the same boundary conditions.
For definiteness, we shall thus demand that at the boundaries of the internal interval}
\beq{}{}{}{}{}{}{}
\psi_\mu \ = \ \ \Lambda\,\psi_\mu \ , \quad \psi_r \ = \ - \ \Lambda\, \psi_r \, \quad \psi_i \ = \ - \,\Lambda\,\psi_i \ , \quad \lambda \ = \ - \Lambda\,\lambda  \ .\label{bcfermi}
\eeq
{ However, the link between the $\Lambda$ eigenvalue of $\lambda$ and the others does not appear compelling for our redefined fields, since eqs.~\eqref{spinor_eqs1_final_text1} do not mix them anymore. Therefore, we shall also explore an additional choice of boundary condition,}
\beq
\lambda \ = \ \Lambda\,\lambda \ . \label{bclambdaplus}
\eeq

Note, finally, that each original ten--dimensional spinor gives rise to a $\underline{4}$ of the $SO(5)$ tangent--space symmetry group of the internal torus. Consequently, the four--dimensional spinor modes that we are about to describe will emerge in quartets.

\subsection{\sc Four--Dimensional Spin--$\frac{3}{2}$ Modes}

We can now begin our analysis, considering the four--dimensional spin--$\frac{3}{2}$ modes arising from the first of eqs.~\eqref{spinor_eqs1_final_text1}. Now $\psi_\mu$, the space--time gravitino component, will be the only non--vanishing field, and will be subject to the constraint
\beq{}{}{}{}{}{}{}{}{}{}{}{}{}{}{}{}{}{}{}{}
\gamma^\mu\,\psi_\mu \ = \ 0 \ . \label{gammatracepsi}
\eeq
We also confine our attention to the $\mathbf{k}=0$ sector, where $\mathbf{k}$ denotes the momentum on the internal torus, since massless fermions can only occur within it, and begin by decomposing the gravitino field according to
\beq
\psi_\mu \ = \ \psi_\mu^+ \ + \ \psi_\m^- \ ,
\eeq
where
\beq
\Lambda\,\psi_\mu^\pm \ = \ \pm \ \psi_\mu^\pm \ .
\eeq
Taking into account that $\Lambda$ anticommutes with $\gamma_\mu$ and commutes with $\gamma_r$, the results collected in Appendix~\ref{app:fermiEinstein32}, and in particular eqs.~\eqref{eqs32}, imply that the spacetime components of the gravitino equation in~\eqref{spinor_eqs1_final_text1} can be cast in the form
\beq
\gamma^{\mu\nu\rho} \ \partial_\nu\,\psi_\rho^\pm \ + \
\gamma_r\,\left[e^{A-B}\left(\partial_r\,+\,A'\,+\,\frac{5}{2}\,C'\right)\,\pm\, {\cal W}_5 \right] \psi^{\mu \,\mp} \ = \ 0  \ , \label{grav_st1}
\eeq
where
\beq
{\cal W}_5 \ = \ \frac{H}{2}\ e^{A-5C} \ , \label{calH5}
\eeq
a concise notation that will recur in the following.

It will be convenient to work in terms of the variable $z$, defined via
\beq
dz \ = \ e^{B-A}\, dr \ , \label{dzdr}
\eeq
with $z(0)=0$. Notice that $z$ has a finite range for the upper branch of $E>0$ solutions in eqs.~\eqref{back_epos_fin2}, or equivalently in eqs.~\eqref{4d_PhiEpos2}, $0\leq z \leq z_m$, with the finite value $z_m$ given by
\beq{}{}{}{}{}{}{}{}{}{}{}{}{}
z_m \ = \ \int\, e^{B-A}\ dr \ \simeq \ 2.24 \, z_0 \ , \label{zm_app}
\eeq
where
\beq{}{}{}{}{}{}{}{}{}{}{}{}{}{}{}{}{}{}{}{}
z_0 \ = \ \left(2\,H \,\rho^3\right)^\frac{1}{2} \ = \ \rho\, h^\frac{1}{2}  \label{z0}
\eeq
was already associated to the supersymmetry breaking scale.
Here and in the rest of the paper $z$-derivatives will often be denoted by a subscript, so that, for instance
\beq{}{}{}{}{}{}{}{}{}{}{}{}{}{}{}{}{}{}{}{}
A_z \ \equiv \ \frac{dA}{dz}\ = \ e^{A-B}\, \frac{dA}{dr}  \ .
\eeq

In this fashion, the preceding spacetime spin--$\frac{3}{2}$ equation becomes
\beq
\gamma^{\mu\nu\rho} \ \partial_\nu\,\psi_\rho^\pm \ + \
\gamma_r\, \left(\partial_z\,+\,A_z\,+\,\frac{5}{2}\,C_z\,\pm\, {\cal W}_5 \right) \psi^{\mu\,\mp} \ = \ 0 \ , \label{grav_st}
\eeq
and its $\gamma$--trace, together with eq.~\eqref{gammatracepsi}, implies that
\beq{}{}{}{}{}{}{}{}{}{}{}{}{}{}{}{}{}{}{}{}
\partial^\mu\,\psi_\mu^\pm \ = \ 0 \ .
\eeq
Moreover, the radial and internal gravitino equations from Appendix~\ref{app:fermiEinstein32} are identically satisfied.

One can separate variables letting
\beq{}{}{}{}{}{}{}{}{}{}{}{}{}{}{}{}{}{}{}{}
\psi_\mu^\pm(x,z) \ = \ \Xi_\mu^\pm(x)\ f^\pm(z) \ , \label{separation}
\eeq
and the spacetime gravitino equation implies that
\beq{}{}{}{}{}{}{}{}{}{}{}{}{}{}{}{}{}{}{}{}
\left(\partial_z\,+\,A_z\,+\,\frac{5}{2}\,C_z\,\pm\, {\cal W}_5 \right)f^\mp \ = \ \alpha^\pm \, f^\pm \ , \label{separation-32}
\eeq
where the $\alpha^\pm$ are a pair of constants, while
\beq
\gamma^{\mu\nu\rho}\,\partial_\nu\,\Xi_\rho^\pm \ = \ \alpha^\pm \, \gamma_r\, \gamma^{\mu\rho}\,\Xi_\rho^\mp \ . \label{gamma-trace}
\eeq
A further step, which will recur in the following, is a redefinition that in this case reads
\beq{}{}{}{}{}{}{}{}{}{}{}{}{}{}{}{}{}{}{}{}
f^\pm \ = \ g^\pm \ e^{\,-\,A\,-\,\frac{5}{2}\,C} \ , \label{fg}
\eeq
which turns the system for $f^\pm$ into the form
\beq{}{}{}{}{}{}{}{}{}{}{}{}{}{}{}{}{}{}{}{}
{\cal A}\, g^- \ = \ \alpha^+\, g^+ \ , \qquad {\cal A}^\dagger\, g^+ \ = \ - \ \alpha^-\, g^- \ ,
\eeq
where
\beq
{\cal A} \ = \ \partial_z \ + \ {\cal W}_5 \ , \qquad {\cal A}^\dagger \ = \ - \ \partial_z \ + \ {\cal W}_5 \ .
\eeq
Notice that the system for $g^\pm$, together with the boundary condition
\beq{}{}{}{}{}{}{}{}{}{}{}{}{}{}{}{}{}{}{}{}
g^+\,g^- \ = \ 0
\eeq
at the ends of the interval, implies that
\beq{}{}{}{}{}{}{}{}{}{}{}{}{}{}{}{}{}{}{}{}
\alpha^+ \int_0^{z_m} \left|g_+\right|^2 \ = \ - \ \alpha^- \int_0^{z_m} \left|g_-\right|^2 \ ,
\eeq
so that either the $\alpha^\pm$ are both zero, or one can set $\alpha^\pm \ = \ \pm\, m$ by a rescaling of $g^\pm$. As a result, in all cases the system can be presented in the manifestly Hermitian form
\beq{}{}{}{}{}{}{}{}{}{}{}{}{}{}{}{}{}{}{}{}
{\cal A}\, g^- \ = \ m\, g^+ \ , \qquad {\cal A}^\dagger\, g^+ \ = \ m \, g^- \ , \label{herm_grav_system}
\eeq
which also leads one to identify the norm
\beq{}{}{}{}{}{}{}{}{}{}{}{}{}{}{}{}{}{}{}{}
\int_0^{z_m} dz \left( \left|g_+\right|^2 \ + \ \left|g_-\right|^2\right) \ .
\eeq
At the same time, the space--time equation becomes
\beq{}{}{}{}{}{}{}{}{}{}{}{}{}{}{}{}{}{}{}{}
\partial \!\!\! / \,\Xi_\mu^\pm(x) \ = \ \pm \, m\, \gamma_r\, \Xi_\mu^\mp(x) \ ,
\eeq
so that $m$ is the four--dimensional mass of these gravitino modes.

Notice that we derived the proper norm for the spin--$\frac{3}{2}$ modes insisting on the reduction to a manifestly Hermitian system. For these modes there is an alternative, more conventional way, of obtaining this result. One can start from the Rarita--Schwinger action contained in eq.~\eqref{action_bergf}, which we write as
\beq{}{}{}{}{}{}{}{}{}{}{}{}{}{}{}{}
{\cal S} \ = \ - \ \frac{1}{k_{10}^2}\ \int d^{10}x \  \sqrt{-g}\  \overline{\psi}_M\, \Gamma^{MNP}\,\partial_N\,\psi_P \ + \ \ldots \ ,
\eeq
in the conventions of Appendix~\ref{app:fermiEinstein}. Separating variables as in eq.~\eqref{separation} leads to
\beq{}{}{}{}{}{}{}{}{}{}{}{}{}{}{}
{\cal S} \ = \ - \ \frac{1}{k_{10}^2}\ {||} f {||}^2\, V_5\, \int d^4\,x \  \overline{\Xi}_\mu \, \gamma^{\mu\nu\rho}\,\partial_\nu\,\Xi_\rho  \ + \ \ldots \ , \label{32action}
\eeq
where $V_5$ denotes the volume of the internal torus.  The norm of $f$ thus induced taking into account all vielbein and metric factors present in the ten--dimensional action reads
\beq{}{}{}{}{}{}{}{}{}{}{}{}{}{}{}{}{}
{||} f {||}^2 \ = \ \int_0^\infty\ dr\ e^{B+4 A+ 5 C - 3 A} \left|f(r)\right|^2 \ . \label{32norm}
\eeq
In terms of $z$ and $g(z)$, which is related to $f$ according to eq.~\eqref{fg}, this result is simply
\beq{}{}{}{}{}{}{}{}{}{}{}{}{}{}{}{}{}
\int_0^{z_m} dz \ \left|g(z)\right|^2 \ ,
\eeq
and there is thus precise agreement between the indications of the ten--dimensional action principle and those drawn from the Schr\"odinger system.

The mass spectrum of spin--$\frac{3}{2}$ modes is fully determined by the normalized solutions of the system of eq.~\eqref{herm_grav_system}, subject to the Fermi boundary conditions~\eqref{bcfermi}, which translate into the demand that $g^-$ vanish at the boundary. General features of this type of system are discussed at length in Appendix~\ref{app:sturmliouville}: for nonzero $m$ it is equivalent to either of the Schr\"odinger--like equations
\beq
{\cal A}\,{\cal A}^\dagger \,g^+ \ = \  m^2 \, g^+ \ , \qquad
{\cal A}^\dagger\,{\cal A}\, g^- \ = \  m^2 \, g^- \ , \label{schrod_gravitino}
\eeq
and the boundary conditions $g^+ \,g^-=0$ at the ends of the interval, which are implied by the Fermi boundary conditions~\eqref{bcfermi}, grant that $m^2 \geq 0$.

\begin{figure}[ht]
\centering
\includegraphics[width=80mm]{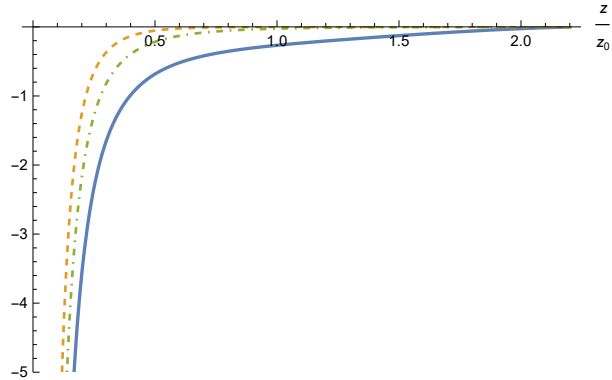}
\caption{\small The potential $V_+$ for the upper branch of $E>0$ gravitino modes (blue, solid), the potential $V_-$ for the lower branch of $E>0$ gravitino modes (orange, dashed) and the supersymmetric potential $V_{\mathrm{susy}}$ (green, dash-dotted). The vertical lines are in units of $\frac{1}{z_0^2}$, where $z_0$ is defined in eq.~\eqref{z0}.}
\label{fig:pot_gG}
\end{figure}
For the upper branch of $E>0$ backgrounds there is a discrete spectrum of normalizable solutions subject to the boundary condition that $g^-$ vanish at the ends of the interval for $z$, consistently with eq.~\eqref{bcfermi}. This is guaranteed by the finite range of the $z$ variable and by the shape of the Schr\"odinger potential for $g^+$,
\bea{}{}{}
V_+ &=& {\cal W}_5^2 \ + \ {\cal W}_5\left(A_z \,-\,5\,C_z\right) \nonumber \\
&=&  \frac{1}{32\,H\,\rho^3}\,\frac{e^{\frac{r}{\rho}\,\sqrt{\frac{5}{2}}}}{\sinh\left(\frac{r}{\rho}\right)^3}\bigg[-6\,\cosh\left(\frac{r}{\rho}\right) \,+\,\sqrt{10}\,\sinh\left(\frac{r}{\rho}\right)\,+\,1\bigg]\ , \label{potential_g+}
\eea
which is displayed in fig.~\ref{fig:pot_gG} in terms of $\frac{r}{\rho}$. This potential is of the form
\beq{}{}{}{}{}{}{}
V \ = \ \frac{1}{z_0^2}\, f_1\left(\frac{r}{\rho}\right) \ = \ \frac{1}{z_0^2}\, f_2\left(\frac{z}{z_0}\right) \ ,
\eeq
so that the discrete mass spectrum in this $\mathbf{k}=0$ sector depends on the parameters via the combination ${z_0}$ in eq.~\eqref{z0}, so that
\beq{}{}{}{}{}{}{}{}{}{}{}{}{}{}{}{}{}{}{}{}{}
\frac{1}{z_0} \ = \ \frac{1}{\ell\, h^\frac{1}{4} }
\eeq
emerges as Kaluza--Klein scale, and thus as the supersymmetry breaking scale, in this class of models.

For the lower branch of $E>0$ solutions the interval has an infinite length and $z_m$ is also infinite, while
\bea{}{}{}
V_- &=& {\cal W}_5^2 \ + \ {\cal W}_5\left(A_z \,-\,5\,C_z\right) \nonumber \\
&=&  \frac{1}{32\,H\,\rho^3}\,\frac{e^{\,-\,\frac{r}{\rho}\,\sqrt{\frac{5}{2}}}}{\sinh\left(\frac{r}{\rho}\right)^3}\bigg[-6\,\cosh\left(\frac{r}{\rho}\right) \,-\,\sqrt{10}\,\sinh\left(\frac{r}{\rho}\right)\,+\,1\bigg]\ , \label{potential_g-}
\eea
Finally, for $E=0$, which is the supersymmetric case as we have seen,  the interval has an infinite length and $z_m$ is again infinite, while the potential can be obtained from the two preceding results in the limit $\rho \to \infty$, and reads
\beq{}{}{}{}{}{}{}{}{}{}{}{}{}
V_{\mathrm{susy}} \ = \ - \ \frac{5}{32\,H\,r^3} \ = \ - \ \frac{5}{36\,z^2} \ . \label{potential_susy}
\eeq
Note in fact that, in this limit
\beq{}{}{}{}{}{}{}{}{}{}{}{}{}{}{}{}{}{}{}{}
{z} \ = \ \frac{2}{3}\,\left(2\,H\,r^3 \right)^\frac{1}{2} \ .
\eeq

The potentials of eqs.~\eqref{potential_g-} and \eqref{potential_susy} clearly result in continuous spectra of excitations. In particular, in the supersymmetric case there is a continuous spectrum of massive modes that can be simply determined solving the Schr\"odinger equation~\eqref{schrod_gravitino}, subject to the boundary condition that $g^-$ vanish at the origin, for which
\beq
g_m^+ \ = \ c^+ \,\sqrt{m z} \ Y_\frac{1}{3}\left(m z\right) \ , \qquad g_m^- \ = \ - \ c^+ \,\sqrt{m z} \ Y_{-\,\frac{2}{3}}\left(m z\right)  \ ,
\eeq
where the $Y$'s are modified Bessel functions. The explicit $m$ dependence grants these modes a standard $\delta$--function normalization.
\begin{figure}[ht]
\centering
\includegraphics[width=70mm]{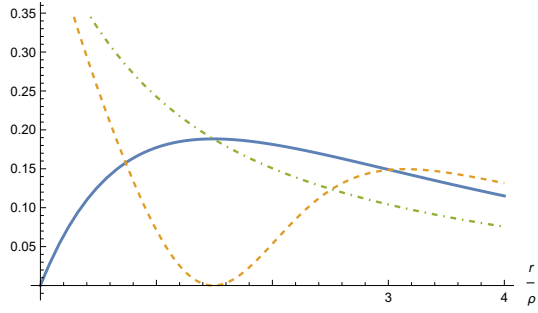}
\caption{ \small The identical $r$--distributions for the ${\psi_\mu}$,  $\lambda^-$ and $\psi_i$ zero modes of eq.~\eqref{gravitino_rdist} (blue, solid), the $r$--distribution for $\xi^-$ of eq.~\eqref{Pixi} (orange, dashed), and the $r$--distribution for $\lambda^+$ of eq.~\eqref{sol_dilatino_plus} (green, dash-dotted) in units of $\frac{1}{\rho}$. The mean values of $r$ are about $4.2\rho$ for the first, $4.4 \rho$ for the second and $3\rho$ for the third.}
\label{fig:distr_Gg}
\end{figure}

If $m=0$, the  equations for $g^+$ and $g^-$ are decoupled and of first order, and the boundary conditions that we have discussed demand that, with supersymmetry not broken by them, $g^-$ should vanish everywhere, so that one is left with
\beq
\left( \partial_z \ - \ {\cal W}_5 \right) g^+  \ = \ 0 \ ,
\eeq
which is solved by
\beq
g^+ \ = \ g_0^+\, e^{\,\int^z\,dz\, {\cal W}_5} \ = \ g_0^+\,e^{\,\frac{H}{2}\int^r \frac{dr}{Y}}\,  \ . \label{gravitino_profile_gen}
\eeq
The integral can be simply computed, and gives
\beq{}{}{}{}{}{}{}{}{}{}{}{}{}
g^+ \ = \
\begin{cases}
    \widetilde{g}_0^+ \, \left[2\,\rho\tanh\left(\frac{r}{2\rho}\right)  \right]^\frac{1}{4} \qquad  &\text{for $E>0$} \ , \\
    \widetilde{g}_0^+ \, {r}^\frac{1}{4} \qquad  &\text{for $E=0$} \ ,
    \label{gravitino_profile}
\end{cases}
\eeq
in terms of a redefined overall constant $ \widetilde{g}_0^+$. The norm for the upper branch of $E>0$ solution is given by
\beq{}{}{}{}{}{}{}{}{}{}{}{}{}{}{}{}{}{}{}{}
\int_0^\infty dr\ e^{B-A}\ \left(g_+\right)^2 \ = \ \int_0^\infty dr\ \left[2\,H\,\rho\, \sinh\left(\frac{r}{\rho}\right)\right]^\frac{1}{2} \ e^{\mp \frac{r \,\sqrt{10}}{4\,\rho}} \  \left(\widetilde{g}_0^+\right)^2 \left[2\,\rho\tanh\left(\frac{r}{2\rho}\right)  \right]^\frac{1}{2} \ ,
\eeq
and it is finite, as we have seen, are the option leading to a finite four--dimensional Planck mass. The other two classes of solutions do not yield normalizable zero modes.

The normalized $r$--distribution for the upper-branch of $E>0$ gravitino zero modes,
\beq
\Pi_{\psi_\mu}\left(\frac{r}{\rho}\right) \ = \ \frac{3}{4\,\rho}\,  \sinh\left(\frac{r}{2\rho}\right)\,e^{\,-\,\frac{r}{2\rho}\,\sqrt{\frac{5}{2}}} \ , \label{gravitino_rdist}
\eeq
is displayed in fig.~\ref{fig:distr_Gg}.

Summarizing, among the upper branch of $E>0$ solutions there is, surprisingly, a quartet of massless symplectic Majorana spin--$\frac{3}{2}$ zero modes whose wavefunctions are normalizable, despite the fact that supersymmetry is broken. Radiative corrections could make these gravitino modes massive if massless spin--$\frac{1}{2}$ goldstini were present. These modes are the subject of the following sections.

\subsection{\sc Spin--$\frac{1}{2}$ Modes from the Ten--Dimensional Gravitino $\psi_M$} \label{sec:tricky_zero_mode}

There are in principle a number of spin--$\frac{1}{2}$ modes of different origin in the backgrounds of Section~\ref{sec:background2}. Let us now focus on modes valued in the spinorial of $SO(5)$, which can be exhibited considering
\beq
\psi_\mu \ = \  \partial_\mu\,\zeta \ + \ \gamma_\mu \chi_1 \ , \qquad \psi_r = \gamma_r\,\chi_2  \ , \qquad \psi_i \ = \ \gamma_i\,\chi_3 \ ,  \label{spinhalfmodes}
\eeq
where $\zeta$, $\chi_1$, $\chi_2$ and $\chi_3$ are spin--$\frac{1}{2}$ fields.
The goldstini are to be found among these modes or others arising from the dilatino, which we shall address in Section~\ref{sec:dilatino}. One can treat them separately since, in our background, the corresponding contributions to eqs.~\eqref{spinor_eqs1_final_text1}  are decoupled.

Making use of the gauge symmetry in the spinor equations collected in Appendix~\ref{app:fermiEinstein12}, one can remove $\zeta$ without affecting $\lambda$. One is thus left with the three $\chi_i$ fields, which  mix in general, as we are about to see.
The results in Appendix~\ref{app:fermiEinstein12}
 determine the three independent components of the gravitino equation, starting from eq.~\eqref{spinor_eqs1_final_text1}. Letting $\phi'=0$ in eq.~\eqref{grav12mu} yields the first of these,
\bea
0&=& e^{-2A}\gamma^{\mu\nu}\,\partial_\nu \left( 2 e^{-A}\,\chi_1 + e^{-B}\,\chi_2 +5 e^{-C}\,\chi_3\right) \nonumber \\ &+&\ \gamma_\mu\,\gamma_r\,e^{-2A}\left[ \frac{3}{2}\,A_z \left( 4 e^{-A}\,\chi_1 - e^{-B}\,\chi_2 + 5\,e^{-C}\,\chi_3 \right) + \partial_z\left( 3\, e^{-A}\chi_1 + 5\,e^{-C}\chi_3\right)\right. \label{st_grav_spinor_text} \\ &+&\ \left.\frac{5}{2}\,C_z\left( 3\, e^{-A}\,\chi_1\,-\,e^{-B}\,\chi_2 \,+\, 6\, e^{-C}\,\chi_3\right) + {\cal W}_5\,\Lambda\left( 3\,e^{-A} \,\chi_1 \ {+} \ e^{-B} \,\chi_2\right)\right]  \, , \nonumber
\eea
which we have recast in terms of $z$-derivatives.

This equation, however, requires some additional comments. It is of the form
\beq
\gamma^{\mu\nu}\,\partial_\nu \Psi_1 \ + \ \gamma^\mu\,\Psi_2 \ = \ 0 \ , \label{eq12_text}
\eeq
with
\bea
\Psi_1 &=& e^{-2A}\left( 2 e^{-A}\,\chi_1 + e^{-B}\,\chi_2 +5 e^{-C}\,\chi_3\right) \ , \nonumber \\
\Psi_2 &=& \gamma_r\,e^{-2A}\bigg[ 3\left( \partial_z \,+\, 2 A_z \,+\, \frac{5}{2}\,C_z \,+\,{\cal W}_5\,\Lambda\right)e^{-A}\,\chi_1\nonumber \\
&-& \frac{1}{2}\Big(3A_z+ 5 C_z\,-\, 2\, {\cal W}_5\,\Lambda\Big)e^{-B}\,\chi_2 \,+\, 5\left(\partial_z + \frac{3}{2}\,A_z+ 3 C_z\right)e^{-C}\,\chi_3  \bigg] \, , \label{psi12_text}
\eea
and we now show that it implies, for the modes of interest, the two equations
\beq
\Psi_1 \ = \ 0 \ , \qquad \Psi_2 \ = \ 0 \ .  \label{psi120}
\eeq
To this end, one can take the divergence and the $\gamma$ trace of eq.~\eqref{eq12_text}, which gives
\beq
\gamma^\mu\,\partial_\mu\,\Psi_2 \ = \ 0 \ , \qquad
 3\, \gamma^{\nu}\,\partial_\nu \Psi_1 \ + \ 4\,\Psi_2 \ = \ 0  \ .
\eeq
Then, making use of these in eq.~\eqref{eq12_text} gives
\beq
\partial_\mu \Psi_1 \ + \ \frac{1}{3}\,\gamma_\mu\,\Psi_2 \ = \ 0 \ , \label{psi122_text}
\eeq
whose curl leads to
\beq
\left(\gamma_\mu\,\partial_\nu \ - \ \gamma_\nu\, \partial_\mu \right) \Psi_2 \ = \ 0 \ .
\eeq
The $\gamma$--trace of this last condition now implies that $\partial_\nu\,\Psi_2=0$, which for the modes we are after is tantamount to $\Psi_2=0$, and finally eq.~\eqref{psi12_text} also implies that $\Psi_1=0$, which proves the asserted result.

Eqs.~\eqref{psi12_text} suggest to introduce the redefined fields
\beq
\tilde{\chi_1} \ = \ e^{-A}\,\chi_1 \ , \qquad \tilde{\chi_2} \ = \ e^{-B}\,\chi_2 \ , \qquad \tilde{\chi_3} \ = \ e^{-C}\,\chi_3 \ .
\eeq
In this fashion, using also eqs.~\eqref{grav12r} and \eqref{grav12i} for the radial and internal Rarita--Schwinger components, the full set of spin--$\frac{1}{2}$ equations becomes
\begin{itemize}
    \item{spacetime:}
    \bea
&& 2\,\tilde{\chi}_1 + \tilde{\chi}_2 +5 \tilde{\chi}_3  = 0 \ ;  \label{alg_chi2} \\
&&  3 \Big(\partial_z+2 A_z + \frac{5}{2}\,C_z \,+ \, {\cal W}_5\,\Lambda \Big)\tilde{\chi}_1\,-\,\frac{1}{2}\Big(3\, A_z\,+\,5\,C_z \,-\, 2\,{\cal W}_5\,\Lambda\Big)\tilde{\chi}_2 \nonumber \\ &&+\ 5\left(\partial_z+\frac{3}{2}\,A_z+ 3\,C_z\right) \tilde{\chi}_3  \ = \ 0 \ ;
\eea
    \item{radial:}
    \bea
&& \Big( 3\,\gamma_r\,\partial \!\!\! / \,  + 6 A_z + 10 C_z \,+\,4\,{\cal W}_5\,\Lambda \Big)\tilde{\chi}_1\,+ \, 5\Big[ \gamma_r\,\partial \!\!\! / \ + 2 \left(A_z+C_z \right) \Big]\tilde{\chi}_3  \  = \ 0 \, ;
\eea
    \item{internal:}
    \bea
&&\left[ 3\,\gamma_r\,\partial \!\!\! / \, +\,  4\left(\partial_z + \frac{5}{2}\,A_z \,+\,2\,C_z\right) \right]\tilde{\chi}_1 \ + \ \left[ \gamma_r\,\partial \!\!\! / \,-\,2\left(A_z+C_z\right)\right] \tilde{\chi}_2 \nonumber \\
&+& 4\Big( \gamma_r\,\partial \!\!\! / \, +\,{\partial_z}+\,{ 2} A_z \,+ \frac{5}{2}\,C_z\,+\, {\cal W}_5\,\Lambda\Big)\tilde{\chi}_3  \ =\ 0 \ .
\eea
\end{itemize}

In order to analyze this system, it is now convenient to eliminate $\tilde{\chi}_2$ using eq.~\eqref{alg_chi2}, while also working in terms of the two combinations
\beq
\Xi_1 \ = \ 3 \,\tilde{\chi}_1 \ + \ 5\, \tilde{\chi}_3 \ , \qquad \Xi_2 \ = \ \tilde{\chi}_1 \ + \ \tilde{\chi}_3  \ . \label{Xichitilde}
\eeq
Taking into account the elimination of $\tilde{\chi}_2$, one can invert these relations, obtaining
\beq{}{}{}{}{}{}{}{}{}{}{}{}
\tilde{\chi}_1 \ = \ -\,\frac{1}{2}\,\Xi_1\,+\,\frac{5}{2}\,\Xi_2 \ , \quad \tilde{\chi}_2 \ = \, -\,\frac{3}{2}\,\Xi_1\,+\,\frac{5}{2}\,\Xi_2 \ , \quad \tilde{\chi}_3 \ = \ \frac{1}{2}\,\Xi_1\,-\,\frac{3}{2}\,\Xi_2 \ ,
\eeq
and one is thus led to the system
\bea
&&\Big(\partial_z + 3 A_z\,+\,\frac{15}{2}\,C_z { \ - \ 3\,{\cal W}_5\,\Lambda}\Big)\Xi_1 \, -\,10 \, \Big(C_z { \,-\, {\cal W}_5\,\Lambda}\Big) \Xi_2 \,=\,0 \ , \nonumber \\
&& \Big(\gamma_r\,\partial \!\!\! / \, + \, 2\,A_z\,-\,2\,{\cal W}_5\,\Lambda\Big)\Xi_1 \, +\, 10 \Big(C_z \,+\, {\cal W}_5\,\Lambda\Big) \Xi_2 \,=\,0  \ , \label{eqsxi123_1}\\
&& \!\!\!\!\! \frac{1}{4}\Big(-\, \gamma_r\,\partial \!\!\! / \,+\,2\,A_z + 4\,C_z \,+\, 2\,{\cal W}_5\,\Lambda\Big)\Xi_1 + \left(\gamma_r\,\partial \!\!\! / + \partial_z + 2\,A_z - \frac{3\,{\cal W}_5}{2}\,\Lambda \right)\Xi_2 \, =  \, 0 \, . \nonumber
\eea
This is actually a set of two coupled systems for the fields $\Xi_i^\pm(x,z)$, where, as in previous cases,
\beq{}{}{}{}{}{}{}
\Lambda \, \Xi_i^\pm(x,z) \ = \ \pm \, \Xi_i^\pm(x,z) \ . \label{lambda_proj}
\eeq
The mass--shell conditions read
\beq
\gamma_r\,\partial \!\!\! / \, \Xi_{i}^\pm \ = \ \mp \, m \,  \Xi_{i}^\mp \ , \label{massshell}
\eeq
and after separating variables, letting
\beq{}{}{}{}{}{}{}{}{}{}{}{}{}{}{}{}{}{}{}{}
\Xi_i^\pm(x,z) \ = \ \Xi_i^\pm(x) \ \xi_i^\pm(z) \label{Xixi}
\eeq
they reduce the systems to
\bea
&&\Big(\partial_z + 3 A_z\,+\,\frac{15}{2}\,C_z  { \ \mp \ 3\,{\cal W}_5} \Big)\xi_1^\pm \, - \, 10\left(C_z \mp \,{\cal W}_5\right) \xi_2^\pm \, = \, 0 \ , \nonumber \\
&& 2\Big(A_z\,\mp\,{\cal W}_5 \Big)\xi_1^\pm \, +\, 10 \Big(C_z \,\pm \,{\cal W}_5\Big) \xi_2^\pm \, = \, \pm\,m\,\xi_1^\mp \ , \label{eqsxi123_22_text}\\
&& \!\!\!\!\! \frac{1}{2}\Big(A_z + 2\,C_z \,\pm \,{\cal W}_5 \Big)\xi_1^\pm + \left(\partial_z\,+\,2\,A_z \mp \frac{3}{2}\,{\cal W}_5 \right)\xi_2^\pm \ =  \ \mp\,m\left(\frac{1}{4}\,\xi_1^\mp\,-\,\xi_2^\mp\right) \ , \nonumber
\eea
while the boundary conditions demand that
\beq
\left.\xi_i^+\right|_{\partial {\cal M}} \ = \ 0 \ ,  \label{bsxii}
\eeq
$(i=1,2)$,
as we have explained at the beginning of Section~\ref{sec:Fermi modes}.

The system consists of six equations for the four unknowns $\xi_1^\pm$ and $\xi_2^\pm$. However, one can verify its consistency taking the derivative of the second equation, which is a pair of algebraic constraints. The derivative vanishes modulo the other equations in~\eqref{eqsxi123_1} and the equations for the background, which in terms of $z$-derivatives become
\bea
&& 3\left(A_z\right)^2 \ + \ 10\, A_z\, C_z \ + \ 5 \left(C_z\right)^2 \ = \ - \ 2\, {\cal W}_5^2 \ , \nonumber \\
&& C_{zz} \ = \ - \ 4\,{\cal W}_5^2 \ - \ \left(3 A_z + 5 C_z\right)C_z\ , \nonumber \\
&& A_{zz} \ = \ 4\,{\cal W}_5^2 \ -  \ \left(3 A_z + 5 C_z\right )A_z \ , \label{hamiltonian_F}
\eea
taking into account the relation between $z$ and $r$ in eq.~\eqref{dzdr}. Before analyzing the system in general, we now consider the supersymmetric limit, which entails some subtleties and deserves a few additional comments.

\subsubsection{\sc The Supersymmetric Limit}

In the supersymmetric limit, which is reached as $\rho \to \infty$,
\beq{}{}{}{}{}{}{}{}{}{}{}{}{}{}{}{}{}{}{}{}
A_z \ = \ - \ C_z \ = \ - \ {\cal W}_5 \ = \ - \ \frac{1}{6\,z} \ ,\label{susylimit}
\eeq
and the complete system reduces to the two sets of equations
\bea
&&\left(\partial_z + \frac{1}{4\,z} \right)\xi_1^+ \, = \, 0 \ , \nonumber \\
&& -\ \frac{2}{3\,z}\,\xi_1^+ \, +\, \frac{10}{3\,z} \, \xi_2^+ \, = \, m\,\xi_1^- \ , \label{eqsxi123_22_text1}\\
&& \frac{1}{6 z}\,\xi_1^+ + \left(\partial_z\,-\,\frac{7}{12\,z} \right)\xi_2^+ \ =  \ -\,m\left(\frac{1}{4}\,\xi_1^-\,-\,\xi_2^-\right) \ , \nonumber
\eea
and
\bea
&&\left(\partial_z + \frac{5}{4\,z} \right)\xi_1^- \, - \,\frac{10}{3\,z} \, \xi_2^- \, = \, 0 \ , \nonumber \\
&& m\,\xi_1^+ \ = \ 0 \ , \label{eqsxi123_22_text2}\\
&& \left(\partial_z- \ \frac{1}{12\,z}\right)\xi_2^- \ =  \ m\left(\frac{1}{4}\,\xi_1^+\,-\,\xi_2^+\right) \ . \nonumber
\eea

For $m \neq 0$, one of the preceding equations demands that $\xi_1^+=0$, so that the system becomes
\bea
&&\frac{10}{3\,z} \, \xi_2^+ \, = \, m\,\xi_1^- \ , \label{eqsxi123_22_text3}\\
&& \left(\partial_z\,-\,\frac{7}{12\,z} \right)\xi_2^+ \ =  \ -\,m\left(\frac{1}{4}\,\xi_1^-\,-\,\xi_2^-\right) \ , \nonumber \\
&&\left(\partial_z + \frac{5}{4\,z} \right)\xi_1^- \, - \,\frac{10}{3\,z} \, \xi_2^- \, = \, 0 \ , \nonumber \\
&& \left(\partial_z \ -\ \frac{1}{12\,z}\right)\xi_2^- \ =  \ - \ m\,\xi_2^+ \ . \nonumber
\eea
One can now link $\xi_1^-$ to $\xi_2^+$, according to
\beq{}{}{}{}{}{}{}{}{}{}{}{}{}{}{}{}{}{}{}{}
\xi_1^- \ = \ \frac{10}{3\,m\,z}\, \xi_2^+\ ,  \label{xi12minus}
\eeq
and the system reduces to the two equations
\beq
 \left(\partial_z \ + \ \frac{1}{4\,z}\right) \xi_2^+ \ = \ m\, \xi_2^- \ , \qquad \left(-\,\partial_z \ + \ \frac{1}{12\,z}\right) \xi_2^- \ = \ m\, \xi_2^+ \ .
\eeq
Redefining the two wavefunctions according to
\beq
\xi_2^\pm \ = \ z^{-\frac{1}{12}}\, \zeta_2^\pm \ ,\label{xi2zeta}
\eeq
now leads to the manifestly Hermitian form
\beq
\left(\partial_z \ + \ \frac{1}{6\,z}\right) \zeta_2^+ \ = \ m\, \zeta_2^- \ , \qquad \left(-\,\partial_z \ + \ \frac{1}{6\,z}\right) \zeta_2^- \ = \ m\, \zeta_2^+ \ ,
\eeq
and identifies the norm
\beq
\int dz \ \left(\left|\zeta_2^+\right|^2 + \left|\zeta_2^-\right|^2 \right) \ .
\eeq
Demanding that $\xi_2^+$ vanish at the origin,
the system is solved by
\beq
 \zeta_2^+ \ = \  C\,\sqrt{m\,z}\,J_{\frac{2}{3}}\left(m z\right) \ , \qquad
  \zeta_2^- \ = \ C\,\sqrt{m\,z}\,J_{-\frac{1}{3}}\left(m z\right)  \ ,
\eeq
where the explicit $m$ dependence grants a conventional $\delta$--function normalization to this continuous spectrum of $\delta$--normalizable wavefunctions. This result is consistent with the expectation that the theory be five dimensional in the limit $\rho \to \infty$. Taking into account eqs.~\eqref{xi12minus} and \eqref{xi2zeta} thus leads to
\bea
&& \xi_2^+ \ = \ C\, \sqrt{m}\, z^\frac{5}{12}\,J_{\frac{2}{3}}\left(m z\right) \ , \qquad \xi_2^- \ = \ C\, \sqrt{m}\, z^\frac{5}{12}\,J_{-\frac{1}{3}}\left(m z\right) \ , \nonumber \\
&& \xi_1^- \ = \ \frac{10\,C}{3\,\sqrt{m}}\ z^{\,-\,\frac{7}{12}}\ J_{\frac{2}{3}}\left(m z\right) \ . \label{cont_spectrum}
\eea
Note that all these wavefunction disappear in the $m \to 0$ limit, since they behave as
\beq
\xi_2^+ \ \sim  \ \frac{C\, m^\frac{7}{6}}{2^\frac{2}{3} \, \Gamma\left(\frac{5}{3}\right)}\, z^\frac{13}{12} \ , \qquad \xi_2^- \ \sim \ \frac{2^\frac{1}{3}\,C\,m^\frac{1}{6}}{\Gamma\left(\frac{2}{3}\right)}\, z^\frac{1}{12}\ , \qquad \xi_1^- \ \sim \ \frac{10\,C\,m^\frac{1}{6}}{3\,2^\frac{2}{3}\,\Gamma\left(\frac{5}{3}\right)}\, z^\frac{1}{12} \ .
\eeq

The $m=0$ case should be treated separately, since eqs.~\eqref{eqsxi123_22_text1} and \eqref{eqsxi123_22_text2} become
\bea
&&\left(\partial_z + \frac{1}{4\,z} \right)\xi_1^+ \, = \, 0 \ , \qquad  -\ \frac{2}{3\,z}\,\xi_1^+ \, +\, \frac{10}{3\,z} \, \xi_2^+ \, = \, 0 \ , \label{eqsxi123_22_text10}\\
&& \frac{1}{6 z}\,\xi_1^+ + \left(\partial_z\,-\,\frac{7}{12\,z} \right)\xi_2^+ \ =  \ 0 \ , \nonumber
\eea
and
\beq
\left(\partial_z + \frac{5}{4\,z} \right)\xi_1^- \, = \, 0 \ , \qquad \left(\partial_z \ - \ \frac{1}{12\,z}\right)\xi_2^- \ =  0 \ .  \label{eqsxi123_22_textsusy}
\eeq
They are solved by
\bea{}{}{}{}{}{}{}{}{}{}{}{}{}{}{}{}{}{}
&&\xi_1^+ \ = \ C_1^+ \, z^{\,-\,\frac{1}{4}} \ , \quad \xi_1^- \ = \ C_1^- \, z^{\,-\,\frac{5}{4}} \ , \quad \xi_2^+ \ = \ \frac{1}{5}\, C_1^+ \, z^{\,-\,\frac{1}{4}} \ , \quad \xi_2^- \ = \ C_2^- \, z^{\,\frac{1}{12}}\ , \label{lim_m01}
\eea
but the boundary condition at $z=0$, which sets to zero $+$ components of the perturbations, demands that $C_1^+=0$. Moreover, the scalar product that emerged above indicates that the wavefunction $\zeta$ corresponding to $C_2^-$, proportional to $z^\frac{1}{6}$, should be rejected, since it would be unbounded at infinity, contrary to what we have seen for massive modes with the same behavior at the origin discussed above. Finally, the solution that emerged for $\xi_1^-$ is not a limit of the preceding ones. One can compute the corresponding norm from the action, as we did for spin--$\frac{3}{2}$ modes in eqs.~\eqref{32action} and \eqref{32norm}, taking into account that when only $\xi_1^-$ is present
\beq
\psi_\mu \ = \ - \ \frac{e^A}{2}\,\gamma_\mu\,\Xi_1^- \ , \qquad \psi_r \ = \ - \ \frac{3\,e^B}{2}\,\gamma_r\,\Xi_1^- \ , \qquad \psi_i \ = \ \frac{e^C}{2}\,\gamma_i\,\Xi_1^- \ .
\eeq
The Rarita--Schwinger action in eq.~\eqref{action_bergf} assigns to this excitation an infinite norm, proportional to
\beq{}{}{}{}{}{}{}{}{}{}{}{}{}{}{}{}{}{}{}{}
\int_0^\infty dr \ e^{2B-A} \ \left|\xi_1^-\right|^2 \ ,
\eeq
so that even this mode should be rejected.
Summarizing, in the supersymmetric limit one obtains a continuous spectrum of excitations, which are all in eqs.~\eqref{cont_spectrum}, and the theory lives effectively in five dimensions.

\subsubsection{\sc The Non--Supersymmetric Case}

We can now turn to the general case, with finite values of $\rho$, and let us begin by considering the system of eqs.~\eqref{eqsxi123_22_text} for $m=0$. Eliminating $\xi_1^-$ from the last two equations, one can obtain an equation for $\xi_2^-$, which takes the form
\beq{}{}{}
\left[\partial_r + 2 A'+\frac{5}{2}\,C'- \frac{H}{2}\,e^{4A} - 5\,\frac{\left(A'+C'\right)\left(C'- \frac{H}{2}\,e^{4A}\right)}{A'+\frac{H}{2}\,e^{4A}}\right]\xi_2^- = 0 \ ,
\eeq
after reverting to the original radial variable $r$.
Now one can use the relations
\beq{}{}{}{}{}{}{}{}{}{}{}{}{}{}{}{}{}{}{}{}
A'\ + \ C'\ = \ - \ \frac{1}{2\sqrt{10}\,\rho} \ , \qquad A'\ + \ \frac{H}{2} \ e^{4 A} \ = \ - \ \frac{1}{4\,\rho} \, \tanh\left(\frac{r}{2\rho}\right) \ ,
\eeq
and then, in the parametrization of eqs.~\eqref{back_epos}, the solutions for $\xi_1$ and $\xi_2$ read
\bea{}{}{}
\xi_2^- &=& e^{\frac{3 r}{8\rho}\sqrt{10}}\left[\sinh\left(\frac{r}{\rho}\right)\right]^{-\frac{1}{8}}\,\left[\tanh\left(\frac{r}{2\rho}\right)\right]^\frac{1}{4}\, \left[\sinh\left(\frac{r}{2\rho}\right)\right]^{-1}\,\xi_{2,0}^- \ , \nonumber \\
\xi_1^- &=& 5\left[1 \ - \ \frac{2}{\sqrt{10}}\,\coth\left(\frac{r}{2\rho}\right)\right]\xi_2^- \ . \label{massless_xi12}
\eea
where $\xi_{2,0}^-$ is a constant. One could solve the system for $\xi_1^+$ and $\xi_2^+$ in a similar fashion, but the corresponding solutions do not vanish at the origin, and must therefore be rejected, since they do not satisfy the boundary conditions discussed at the beginning of Section~\ref{sec:Fermi modes}.

Note that the solutions that we have displayed have no singularities in the interior of the $r$-interval, but they are singular at $r=0$ and at $r=+\infty$. Furthermore, $\xi_1^-$ vanishes in the interior of the interval, at $r=r_0$, where
\beq
\tanh\left(\frac{r_0}{2\rho}\right) \ = \ \frac{2}{\sqrt{10}} \ , \label{r0}
\eeq
which corresponds to $r_0 \approx 1.49 \rho$. We shall shortly have more to say about this special point, but let us also note, for the time being, that up to a proportionality constant
\beq
\xi_1^- = e^{\frac{3 r}{8\rho}\sqrt{10}}\left[\sinh\left(\frac{r}{\rho}\right)\right]^{-\frac{1}{8}}\,\left[\tanh\left(\frac{r}{2\rho}\right)\right]^\frac{1}{4}\, \left[\sinh\left(\frac{r}{2\rho}\right)\right]^{-2} \sinh\left(\frac{r-r_0}{2\rho}\right) \ . \label{Xi1m}
\eeq

Although we have presented a simple derivation of these zero modes, we cannot make any definite statement about their normalizability yet. As in previous cases, the proper setup for the massive spectrum will determine the precise form of the normalization integrals. This is particularly convenient for this set of modes, since their mixing makes it less handy to deduce this result directly from the low--energy effective action.

In the general massive case, the structure of the system of eqs.~\eqref{eqsxi123_22_text} suggests to multiply the first equation by $C_z \pm {\cal W}_5$, the second by $C_z \mp {\cal W}_5$ and add them, while also defining
\beq{}{}{}{}{}{}{}{}{}{}{}{}{}{}{}{}{}{}{}{}
\widehat{\xi}^\pm \ = \ \Big(C_z \ \pm \ {\cal W}_5\Big)\,\xi_1^\pm \ .
\eeq
These steps lead to the two coupled equations for $\widehat{\xi}_1^\pm$
\bea
&&\Big(\partial_z + 3 A_z\,+\,\frac{15}{2}\,C_z \,\mp\, 3\,{\cal W}_5\Big)\,\widehat{\xi}^\pm \nonumber \\
&&+\ \frac{6\,{\cal W}_5^2 \ + \ 5\,C_z\left(A_z+C_z\right) \, \mp \, 3\,{\cal W}_5\left(A_z-C_z\right)}{C_z \,\pm\,{\cal W}_5} \,\widehat{\xi}^\pm  \, = \, \pm\, m \,\widehat{\xi}^\mp  \label{sisxi1}
\eea
and $\xi_2^\pm$ can then be determined algebraically from $\xi_1^\pm$ via the second of the second of eqs.~\eqref{eqsxi123_22_text}. The resulting system of eqs.~\eqref{sisxi1} is of the form
\bea
&& \left(-\, \partial_z \ + \ \Omega^+ \right) \widehat{\xi}^+ \ = \ m \, \widehat{\xi}^-  \ , \nonumber \\
&& \left(\partial_z \ + \ \Omega^- \right) \widehat{\xi}^- \ = \  m \, \widehat{\xi}^+ \ , \label{sysxipm}
\eea
where
\beq
\mp\ \Omega^\pm \ = \ 3 A_z\,+\,\frac{15}{2}\,C_z \,\mp\,{3}\,{\cal W}_5 \ + \ \frac{6\,{\cal W}_5^2 \ + \ 5\,C_z\left(A_z+C_z\right) \ \mp \ 3\,{\cal W}_5\left(A_z-C_z\right)}{\left(C_z\,\pm\,{\cal W}_5\right)} \ .
\eeq

Note that this step has introduced a spurious singularity in the interior of the $r$-interval, where the denominator of $\Omega^-$,
\beq
C_z\,-\,{\cal W}_5 \ = \ \frac{1}{4\rho}\,e^{A-B} \left( \frac{\cosh\left(\frac{r}{\rho}\right)\,-\,1}{\sinh\left(\frac{r}{\rho}\right)}\,-\,\frac{2}{\sqrt{10}} \right)\ ,
\eeq
vanishes. This occurs precisely at the point $r=r_0$ that we already identified in eq.~\eqref{r0}, but the background geometry is not singular there.
In fact, the direct derivation of the massless modes via ${\xi}_2^-$ in eqs.~\eqref{massless_xi12} exhibited no singularity of this type, but interestingly the zero--mode solution ${\xi}_1^-$ in eq.~\eqref{massless_xi12} vanishes precisely at $r_0$. In the following, one should work in principle within regions that do not include this singular point, and then glue together the solutions thus obtained, while also subjecting the result to proper boundary conditions at $r=0$ and at $r=\infty$, which demand that $\xi_1^+$ vanish there.

The system of eqs.~\eqref{sysxipm} is still not in the form we are after, but the further redefinitions
\beq
\widehat{\xi}^\pm(z)\ = \ e^{\,\frac{1}{2}\,\int\left(\Omega^+ \,-\,\Omega^-\right)\,d z}\ \widetilde{\xi}^\pm(z) \
\eeq
lead to the manifestly Hermitian system
\beq
{\cal A}\, {\widetilde{\xi}}^+ \ = \ m \, {\widetilde{\xi}}^-  \ , \qquad {\cal A}^\dagger\,{\widetilde{\xi}}^- \ = \ m \, {\widetilde{\xi}}^+ \ , \label{sysxipmh}
\eeq
with
\beq
{\cal A}^\dagger \ = \ - \ \partial_z \ - \  \frac{\Omega^+\,+\,\Omega^-}{2}  \ , \qquad {\cal A} \ = \  \partial_z \ - \  \frac{\Omega^+\,+\,\Omega^-}{2} \ . \label{aadagxi}
\eeq
Notice that the relevant combination of $\Omega^\pm$ has a simple form:
\beq{}{}{}{}{}{}{}
\frac{\Omega^+ \, + \, \Omega^-}{2} \,=\, - \, {\cal W}_5 \ \frac{5\,C_z \ + \ 9\, A_z}{7\, C_z \ + \ 3\, A_z} \,=\, - \ \frac{e^{A-B}}{4\ \rho} \ \sinh\left(\frac{r_0}{\rho}\right)\ \frac{\coth\left(\frac{r}{\rho}\right)\ + \ \frac{5}{2\sqrt{10}}}{\sinh\left(\frac{r-r_0}{\rho}\right)} \ . \label{opm}
\eeq

This expression has a singularity, a simple pole at the point $r=r_0$ of eq.~\eqref{r0}, which lies inside the interval where, as we have stressed, the background is not singular, and in its neighborhood
\beq
\frac{\Omega^+ \ + \ \Omega^-}{2} \ \sim \ - \ e^{A(r_0)-B(r_0)} \ \frac{1}{r \ - \ r_0} \ .
\eeq
The operators $A$ and $A^\dagger$ inherit singularities at $z=z_s=z(r_0)$, and taking into account that
\beq{}{}{}{}{}{}{}{}{}{}{}{}{}{}{}{}{}{}{}{}
z \ - \ z_s \ \sim \left.\frac{dz}{dr}\right|_{r_0} \left(r \ - \ r_0\right) \ ,
\eeq
and finds
\beq
{\cal A}^\dagger \ \sim \ - \ \partial_z \,+\, \frac{1}{z-z_s} \ , \qquad {\cal A} \ \sim \ \partial_z \,+\, \frac{1}{z-z_s} \ .
\eeq
On the other hand, the Schr\"odinger operator ${\cal A}\,{\cal A}^\dagger$ associated to $\widetilde{\xi}^-$ is not singular, and indeed near $z=z_s$
\beq
{\cal A}\,{\cal A}^\dagger \ \sim \ - \ \partial_z^2 \ .
\eeq

In this fermionic problem, one is confronted again with the type of system discussed in detail in Appendix~\ref{app:sturmliouville}, which implies that the norm should be defined as
\beq{}{}{}{}{}{}{}{}{}{}{}{}{}{}{}{}{}{}{}{}
\int dz \left(\left|\widetilde{\xi}^+\right|^2 \ + \ \left|\widetilde{\xi}^-\right|^2 \right) \ .
\eeq
There we show that, if the product $\widetilde{\xi}^+\,\widetilde{\xi}^-$ vanishes at the boundaries (which here include the point $z=z_s$), the structure of the system grants that $m^2 \geq 0$. In our problem, the $\Lambda$ projection demands indeed that at the ends of the interval, $r=0$ and $r=+\infty$,
\beq{}{}{}{}{}
\widetilde{\xi}^{+} \ = \ 0 \ ,
\eeq
but the spurious singularity demands, in addition, that for $m \neq 0$ $\widetilde{\xi}^-$ vanish at $z=z_s$.  Otherwise $\widetilde{\xi}^+$, which is obtained from it via eq.~\eqref{sysxipmh}, would have a pole there, and consequently a divergent norm.

The massive Fermi spectrum is thus determined by the Schr\"odinger--like equation
\beq
{\cal A}\,{\cal A}^\dagger \, \widetilde{\xi}^- \ = \ m^2\, \widetilde{\xi}^- \ , \label{schrod_xi}
\eeq
to be supplemented with the boundary conditions
\beq{}{}{}{}{}
\lim_{z \to 0} {\cal A}^\dagger \, \widetilde{\xi}^{-} \ = \ 0  \qquad  \lim_{z \to z_s} \widetilde{\xi}^{-} \ = \ 0 \ . \label{bc_xim}
\eeq

Close to the origin, eqs.~\eqref{aadagxi} and \eqref{opm} determine the limiting behavior
\beq{}{}{}{}{}{}{}
{\cal A}^\dagger \ \sim \ - \, \partial_z \ - \ \frac{1}{6\,z} \ ,
\eeq
so that one is demanding that, as $r$ or $z$ tend to zero,
\beq{}{}{}{}{}{}{}
\widetilde{\xi}^- \ \sim \ z^{-\frac{1}{6}} \ \sim \ r^{-\frac{1}{4}} \ . \label{behxim}
\eeq
As we have already mentioned, these boundary conditions grant that, for the problem at stake, $m^2 \geq 0$, but only if they are supplemented by the condition that $\widetilde{\xi}^-$ vanish at the spurious singularity, as pertains to the Fermi system.

We can now take a closer look at the zero modes.
The system of eqs.~\eqref{sysxipm} admits in principle two types of zero modes, in each of the two regions $r<r_0$ and $r>r_0$ obtained leaving out the spurious singularity,
\bea{}{}
\widetilde{\xi}^+ &=& \tanh\left(\frac{r}{\rho}\right)^\frac{1}{4}\coth\left(\frac{r-r_0}{\rho}\right) \left[\zeta_{1}\,\theta(r_0-r)\ +\ \zeta_{2}\,\theta(r-r_0)\right] \ , \nonumber \\
\widetilde{\xi}^-&=& \tanh\left(\frac{r}{\rho}\right)^{-\frac{1}{4}}\tanh\left(\frac{r-r_0}{\rho}\right) \zeta_{0}  \ , \label{xipmshape}
\eea
where $\theta$ denotes the Heaviside step function and $\zeta_{0}$, ${\zeta}_{1}$ and ${\zeta}_{2}$ are constants. The first solution is singular at $r=r_0$ and not normalizable, for any choice of the constants, while the second is regular at $r_0$ and is singular at the origin but is normalizable, since
\beq{}{}{}{}{}{}{}
\int_0^{z_m} dz \ \left|{\widetilde{\xi}}^-\right|^2 \ = \ \left|\zeta_0\right|^2\int_0^\infty dr\  e^{B-A}\ \tanh\left(\frac{r}{\rho}\right)^{-\frac{1}{2}}\,\tanh\left(\frac{r-r_0}{\rho}\right)^2  \label{normintxi}
\eeq
is finite, while $\widetilde{\xi}^+$ vanishes for the proper zero mode.
Moreover, ${\widetilde{\xi}}^-$ is continuous at $r=r_0$, so that it does solve eq.~\eqref{bc_xim} for $r\geq 0$.

\begin{figure}[ht]
\centering
\includegraphics[width=80mm]{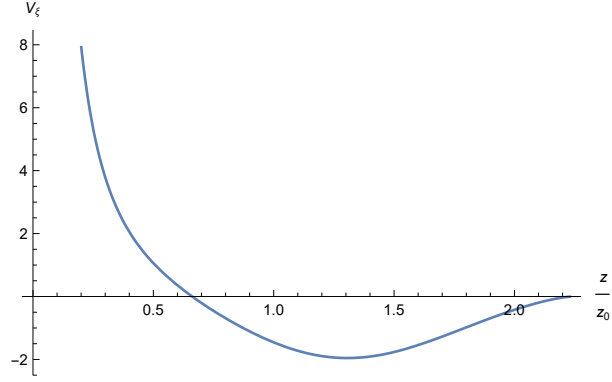}
\caption{\small The potential $V_\xi$ of eq.~\eqref{Vxi}, in units of $\frac{1}{z_0^2}$ and as a function of $\frac{z}{z_0}$, where $z_0$ is defined in eq.~\eqref{z0}. The plots are displayed up to the right end of the interval, $z \,\simeq\, 2.24\,z_0$.}
\label{fig:pot_xi}
\end{figure}

There is an apparent puzzle here, since $\widetilde{\xi}^-$ has a node at $r=r_0$. Consequently, it ought to be the first excited state, rather than the ground state, of the Schr\"odinger problem~\eqref{schrod_xi}, which should therefore have a tachyonic mode. The puzzle is resolved by taking a closer look at the limiting behavior at the spurious singularity $z_s$.
The Schr\"odinger potential for $\widetilde{\xi}^-$ of fig.~\ref{fig:pot_xi},
\bea
{V}_{{\xi}}(r) \!\!\!&=& \frac{1}{4}\left(\Omega^+ \,+\, \Omega^-\right)^2 \  - \ \frac{1}{2}\,\partial_z\left(\Omega^+ \,+\, \Omega^-\right) \nonumber \\
&=& \!\!\!\!-\,\frac{ \left[\left(\cosh\left(\frac{r}{\rho}\right)-1\right)\left(10 \cosh\left(\frac{r}{\rho}\right)-15+7\sqrt{10}\sinh\left(\frac{r}{\rho}\right)\right)-70 \right]e^{\frac{r}{\rho}\sqrt{\frac{5}{2}}}}{192 \, H\,\rho^3\left[\cosh\left(\frac{r-r_0}{2\rho}\right)\right]^{2}\, \sinh\left(\frac{r}{\rho}\right)^3}  \label{Vxi}
\ ,
\eea
is manifestly regular at $r_0$.
It does have a ground state with $m^2<0$, as can be anticipated, for example, resorting to a quadratic approximation around its minimum.
However, \emph{the ground--state wavefunction of the Schr\"odinger problem is not physically acceptable} for the Fermi system~\eqref{sysxipm}, whose boundary conditions~\eqref{bc_xim} select solutions with $A\,\widetilde{\xi}^-$ vanishing at zero and $\widetilde{\xi}^-$ vanishing at $z_s$. The ground--state wavefunction of the Schr\"odinger problem can not vanish at $z_s$, since it has no nodes, so that it does not obey the proper Fermi boundary conditions. Equivalently, the formal positivity argument for ${\cal A}\,{\cal A}^\dagger$ fails unless $\widetilde{\xi}^-$ vanishes at $z=z_s$.

To reiterate, the solutions collected in eqs.~\eqref{massless_xi12} are true normalizable zero modes, and the corresponding normalized $r$--distribution,
\beq{}{}
\Pi_{\widetilde{\xi}} \ \simeq \ \frac{0.61}{\rho}\,e^{\,-\,\frac{r}{2\rho}\,\sqrt{\frac{5}{2}}}\, \left[\cosh\left(\frac{r}{\rho}\right)\right]^\frac{1}{2}\,\left[\tanh\left(\frac{r-r_0}{\rho}\right)\right]^2 \ , \label{Pixi}
\eeq
determined by eq.~\eqref{normintxi}, is displayed in fig.~\ref{fig:distr_Gg}.

Comparing the $r$--dependence in eqs.~\eqref{massless_xi12} and in the second of eqs.~\eqref{xipmshape}, one can now see that, up to a proportionality constant, the measure for ${\xi}_1^-$ is
\beq
\int dr \  \left|{\xi}_1^-\right|^2  e^{\,-\,\frac{r\sqrt{10}}{\rho}}  \left[\frac{\cosh\left(\frac{r-r_0}{2\,\rho}\right)}{\cosh\left(\frac{r-r_0}{\rho}\right)}\right]^2 {\left[\cosh\left(\frac{r}{\rho}\right)\right]^\frac{1}{2}\left[\sinh\left(\frac{r}{\rho}\right)\right]^\frac{3}{4}}\ \left[\sinh\left(\frac{r}{2\,\rho}\right)\right]^3.
\eeq
This intricate measure reflects the presence of the algebraic constraint relating $\chi_2$ to $\chi_1$ and $\chi_3$.

In conclusion, there is a quartet of four--dimensional massless Weyl spinor modes from this sector, which are potential goldstini. The actual goldstini are combinations of these modes with others that can arise from $\lambda$, to which we now turn.

\subsection{\sc Spin--$\frac{1}{2}$ Modes from the Ten--Dimensional Dilatino $\lambda$} \label{sec:dilatino}

The relevant equation for these modes is the second of eqs.~\eqref{spinor_eqs1_final_text1}, which becomes
\bea
&& \left(\gamma_r\,\partial \!\!\! / \,+ \,\partial_z + 2 A_z+\frac{5}{2}\,C_z \right)\lambda(x,z) \  + \ {\cal W}_5 \ \Lambda\,\lambda(x,z)  \ = \ 0  \ .
\eea
Here $\lambda$ is to be decomposed into eigenstates of the matrix $\Lambda$ of eq.~\eqref{lambdamat},
and defining as in the preceding cases the four--dimensional mass via
\beq
\gamma_r\,\partial \!\!\! / \, \lambda^\pm(x,z) \ = \ \mp \, m \, \lambda^\mp(x,z) \ ,
\eeq
leads to
\bea
 && \Big(\partial_z + 2 A_z+\frac{5}{2}\,C_z \ \pm \ {\cal W}_5 \Big)\lambda^\pm(x,z)  \ = \ \pm \, m\, \lambda^\mp(x,z)  \ . \label{syslambda}
\eea
Performing the separation of variables and the additional redefinition
\beq
\lambda^\pm(x,z) \ = \ e^{-2A-\frac{5}{2}\,C} \ h^\pm(z) \,\lambda^\pm(x)
\eeq
yields for these modes the manifestly Hermitian system
\beq
 \Big(\pm\,\partial_z \ + \ {\cal W}_5\Big)h^\pm  \ = \ m\, h^\mp  \ , \label{syslambda2}
\eeq
so that the two $h^\pm$ wavefunctions obey the same equations as the two $g^\mp$ in eqs.~\eqref{herm_grav_system}, and can be identified with them. This is also consistent with the boundary conditions~\eqref{bcfermi}, which now remove $h^+$ at the ends, while in the spin--$\frac{3}{2}$ case they removed $g^-$. As a result, there is a one--to--one correspondence between the massive spin--$\frac{3}{2}$ spectrum arising from the gravitino $\psi_\mu$ and the massive spin--$\frac{1}{2}$ spectrum arising from $\lambda$ with these boundary conditions.

For massless modes the two equations in~\eqref{syslambda2} decouple,
and the proper choice for a normalizable zero mode is
\beq
\lambda^-(x,z) \ = \ e^{-2A-\frac{5}{2}\,C} \left[\tanh\left(\frac{r}{2\,\rho}\right) \right]^{\frac{1}{4}} \, \lambda^-(x) \ . \label{sol_dilatino}
\eeq
{ The alternative boundary condition of eq.~\eqref{bclambdaplus} leads to a different normalizable wavefunction,}
\beq
\lambda^+(x,z) \ = \ e^{-2A-\frac{5}{2}\,C} \left[\tanh\left(\frac{r}{2\,\rho}\right) \right]^{\,-\,\frac{1}{4}} \, \lambda^+(x) \ . \label{sol_dilatino_plus}
\eeq

{  With $\lambda^-$, this sector would thus contribute four massless Weyl spin--$\frac{1}{2}$ modes, which have the quantum numbers of the expected goldstini, together with a discrete spectrum of massive excitations. The $\lambda^-$ $r$--distribution is identical to the one in eq.~\eqref{gravitino_rdist}. In contrast, the $\lambda^+$ distribution is different, and does not vanish as $r \to 0$. With this choice, the goldstini must necessarily originate from section~\ref{sec:tricky_zero_mode}. The dilatino distributions $\lambda^\pm$ are displayed in fig.~\ref{fig:distr_Gg}.}

\subsection{\sc Additional Spin--$\frac{1}{2}$ Modes from $\psi_i$}

The remaining spin--$\frac{1}{2}$ modes can be obtained setting $\psi_\mu=0$, $\psi_r=0$, $\lambda=0$ and $\gamma^i\,\psi_i=0$. As before, we concentrate on modes with $\mathbf{k}=0$, and making use of the results collected in Appendix~\ref{app:fermimodes12i}, one can see that the resulting equation reads
\beq
\left(\partial_z \,+\, \gamma_r\,\partial \!\!\! / \, + \, 2\,A_z\,+\,\frac{3}{2}\,C_z\,+\, {\cal W}_5\,\Lambda \right)\psi_k \ = \ 0 \ . \label{eqpsik}
\eeq
As for the other sectors, one can expand in eigenstates of $\Lambda$ with mass--shell conditions as in eq.~\eqref{massshell}, which yields the system
\beq
\left(\partial_z \,+ \, 2\,A_z\,+\,\frac{3}{2}\,C_z\,\pm\, {\cal W}_5 \right)\psi_k^\pm\ = \ \pm \ m \, \psi_k^\mp \ . \label{eqpsik2}
\eeq
Separating variables according to
\beq
\psi_k^\pm(x,z) \ = \  \psi_k^\pm(x) \ \chi_k^\pm(z)  \ = \ e^{-2A-\frac{3}{2}\,C} \, \psi_k^\pm(x) \ \widetilde{\chi}_k^\pm(z)  \ , \label{psikpm}
\eeq
the system can be turned into the manifestly Hermitian form
\beq
\left( \pm\,\partial_z \ + \ {\cal W}_5\right) \widetilde{\chi}_k^\pm(z) \ = \ m\, \widetilde{\chi}_k^\mp(z) \ , \label{syschipm}
\eeq
so that the corresponding norm is determined by
\beq{}{}
\int dz \ {\widetilde{\chi}_k}{}^\dagger(z)\, \widetilde{\chi}_k(z) \ . \label{normchik}
\eeq
Note that the corresponding $r$-measure of $\psi_k(x,z)$,
\beq{}{}
\int dr \ e^{2B-A-2C}\  {\chi_k}{}^\dagger(z)\, \chi_k(z) \ , \label{normpsik}
\eeq
is precisely as implied by the Rarita--Schwinger kinetic term, taking into account the presence of two upper internal $\Gamma$-matrices and un upper spacetime $\Gamma$-matrix for this set of modes. The system in eq.~\eqref{syschipm} has the same structure as those in eqs.~\eqref{herm_grav_system} and~\eqref{syslambda2}, and therefore both the allowed masses and the resulting distribution of massive modes are identical to what we found for the spin--$\frac{3}{2}$ spectrum.

The boundary condition~\eqref{bcfermi} demands that $\widetilde{\chi}_k^+(z)$ vanish at the ends, thus removing it all together for massless modes, and consequently
\beq{}{}{}{}{}{}{}{}{}{}{}{}{}
\widetilde{\chi}_k^-(z) \ = \widetilde{\chi}_{(0)\,k}^- \, \left[\tanh\left(\frac{r}{2\rho}\right)  \right]^\frac{1}{4} \ ,
    \label{psik_profile}
\eeq
which is normalizable, as was the case for the corresponding spin--$\frac{3}{2}$ modes.

Summarizing, for the upper branch of the $E>0$ solutions, four more quartets of massless Weyl spinor modes emerge from the five $\psi_i$, subject to $\gamma^i\,\psi_i=0$, together with a discrete spectrum of massive modes. The distribution of these modes is determined by the same potentials that already emerged for massive gravitino and dilatino modes.
\section{\sc Conclusions} \label{sec:conclusions}

In this paper we have explored in detail a class of Randall--Sundrum--like~\cite{randallsundrum} compactifications of the type--IIB string to four--dimensional Minkowski space that generically break all supersymmetries. The solutions are supported by a flux of the self--dual five form field strength that is homogeneous in an internal five--torus, and combine an internal interval of finite length with a warped four--dimensional Minkowski spacetime. After a detailed discussion of the properties of these backgrounds,  which completes the results in~\cite{ms21_1} since we have shown that a constant dilaton profile is uniquely selected when taking boundary conditions into account, we have analysed the massless Fermi modes present in them. To this end, we have classified the perturbations according to the infinitesimal global symmetries of four--dimensional Minkowski space and of the internal torus, which allow one to deal separately with different mode sectors.

The nature of the zero modes for Fermi fields that we found { enforcing identical $\Lambda$ projections at the ends of the interval} is summarized in Table~\ref{table:tab_F}.  In all cases, the $\Lambda$ projection introduced by the boundary conditions removes one half of the original Fermi modes, and the reader will not fail to notice that the resulting massless spectrum is that of $N=4$ supergravity coupled to five vector multiplets. The massless Fermi modes originating from $\psi_\mu$, $\lambda$  and $\psi_i$ have identical distributions along the internal interval, which are given in eq.~\eqref{gravitino_rdist}. Moreover, their massive spectra are in one-to-one correspondence and are simple to analyze. On the other hand, the modes associated in the table to $\psi_M$ arise from different components of the ten--dimensional gravitini. The massless modes in this sector have the different radial distribution in eq.~\eqref{Pixi}, and the corresponding massive spectrum is also different. The analysis entails a number of subtleties that are explained in detail
in Section~\ref{sec:tricky_zero_mode}.
\begin{table}[h!]
\centering
\begin{tabular}{||c c c c c ||}
 \hline
$4D\,\mathrm{hel.} \times SO(5)$ & 4D Content &$r$-dist & $10D$ origin & $\Lambda$ \\ [0.5ex]
 \hline\hline
$\left(\pm\,\frac{3}{2},4\right)$ & 4 gravitinos & \eqref{gravitino_rdist}  & $\psi_{\mu}$  & $+ 1$ \\
$\left(\pm\,\frac{1}{2},4\right)$ & 4 spinors & \eqref{gravitino_rdist}   & $\lambda$  & $- 1$  \\
$\left[ \right.$ $\left(\pm\,\frac{1}{2},4\right)$ & 4 spinors & \eqref{sol_dilatino_plus}   & $\lambda$  & $ 1$ $\left. \right]$  \\
{$\left(\pm\,\frac{1}{2},4\right)$} & 4 spinors & \eqref{Pixi}  & $\psi_M$  & $- 1$  \\
$\left(\pm\,\frac{1}{2},4\right)$ & $4 \times 4$ spinors & \eqref{gravitino_rdist} & $\psi_i$  & $- 1$  
 \\ [1ex]
 \hline
\end{tabular}
\caption{ Four--Dimensional massless Fermi Modes. $\Lambda$ denotes the eigenvalue that characterizes identical Fermi boundary conditions at the ends of the interval, as discussed in Section~\ref{sec:Fermi modes}. The alternative choice for the dilatino, with $\Lambda=1$, corresponds to the line within square brackets.}
\label{table:tab_F}
\end{table}

In order to decide whether or not a mode is normalizable, we have cast the equations for the mode profiles within the different sectors into Schr\"odinger--like forms, combining redefinitions of the wavefunctions with a convenient choice for the independent variable. More precisely, we have cast them into fermionic counterparts of Schr\"odinger--like formulations, the first-order systems described in detail in Appendix~\ref{app:sturmliouville}. The available mass spectra were thus determined by the eigenvalues of operators that are Hermitian with respect to the usual $L^2$ scalar product, when they are combined with the boundary conditions discussed in Appendix~\ref{app:fermi_interval}. { More precisely, we focused on the choice of identical $\Lambda$ eigenvalues, for all Fermi fields, at the two ends of the interval. This has the virtue of allowing massless Fermi modes, while choosing opposite eigenvalues at the two ends would eliminate them.} 

Whenever the Schr\"odinger systems are formally self--adjoint, the no--flow conditions for the translation generators in spacetime and along the internal torus, together with the corresponding conditions on the Lorentz generators for Fermi fields, guarantee real mass spectra, and the resulting norms were also instrumental to identify the actual massless modes. In most cases, we also verified that the low--energy field theory yields precisely the very same norms  obtained arising from the Schr\"odinger systems.

For Fermi fields we have
largely focused on massless spectra, since their massive modes cannot be the source of instabilities. We have found a surprising { option} for the massless fermionic spectrum: although supersymmetry is broken in the resulting four--dimensional Minkowski space, there are zero modes originating from half of the original ten--dimensional spin--$\frac{3}{2}$ gravitini and from half of the original ten--dimensional dilatini $\lambda$. { With identical $\Lambda$ projections at the two ends of the interval, the} fermionic zero modes are indeed those of $N=4$ four--dimensional supergravity coupled to \emph{five} vector multiplets, which is surely surprising in the presence of broken supersymmetry but resonates with the presence of an internal $T^5$. Since supersymmetry is fully broken for finite values of $\rho$ or $\ell$, the presence of four--dimensional massless gravitini was clearly unexpected. However, nothing prevents them from acquiring masses, once radiative corrections are taken into account, by absorbing the massless spin--$\frac{1}{2}$ modes that are also present and can mix with them. For example, the massless spectrum includes spin--$\frac{1}{2}$ massless modes arising from the ten--dimensional dilatino, with which they could combine into massive spin--$\frac{3}{2}$ particles. A naive estimate of the resulting mass scale, obtained taking into account the IIB string--scale cutoff, leads on dimensional grounds to
\beq
\Delta m \ \sim \ \frac{1}{m_{Pl(4)}^2}\, \left({M_s}\,h^\frac{1}{4} \right)^3 \ \sim \ \frac{h^\frac{5}{4}\,g_s^2}{M_s^5\,\ell^2\,\Phi} \ = \ \frac{h^\frac{7}{4}\,g_s^2}{M_s^5\,\Phi}\ \mu_S^2 \ . \eeq
Indeed, the string--induced cutoff for four--dimensional amplitudes is the string scale $M_s$, and consequently, taking into account the metric of eq~\eqref{4d_PhiEpos}, the four--dimensional one is ${M_s}\,h^\frac{1}{4}$. Here we have also used eqs.~\eqref{planck4} and \eqref{ell h}, and $m_{Pl(4)}$ is the four--dimensional Planck mass. Altogether, one can conclude that
\beq{}{}{}{}{}{}{}{}{}{}{}{}{}{}{}{}{}{}{}{}{}
\Delta\,m \ \ll \ M_s\,h^\frac{5}{4}\,g_s^2 \ , 
\eeq
which leaves an interesting range of values within the region of validity of the effective field theory identified in Section~\ref{sec:background2}.
Half of the original ten--dimensional supersymmetry is recovered in the limit of large values for $\rho$ or $\ell$, while keeping $\frac{h^\frac{5}{4}}{\ell}$ finite, as discussed in Section~\ref{sec:background2}, so that $\Delta\,m\to 0$. However, the effective field theory ought to be examined with reference to~\cite{cremmer} and the following literature, since this limiting behavior takes place within a curved five--dimensional setting~\footnote{As we have stressed, a second option that emerged from our discussion would have left no massless modes altogether. It consists in implementing the Scherk-Schwarz~\cite{scherkschwarz} mechanism, along the lines of conventional Kaluza--Klein compactifications, by enforcing different boundary conditions at the two ends of the interval.}.

The Schr\"odinger potentials governing the massive Fermi spectra from $\psi_\mu{}^+$, $\lambda^-$ and $\psi_i{}^-$ are identical, but the massive spectrum from the singlet spin--$\frac{1}{2}$ sector arising from $\psi_M$ is different. An amusing technical subtlety actually emerged from the singlet $\psi_M$ sector: the fermionic boundary conditions remove the node-free tachyonic ground state of the associated second--order Schr\"odinger--like equation, so that the actual massless spinor profile has, surprisingly, one node within the $r$-interval. { We have also considered a second option, which results in a massless $\lambda^+$ profile and appears viable for our redefined fields.  

 All the different options that we have explored in detail or clearly addressed, including those where all Fermi modes are massive, appear viable at this stage. There is apparently some tension between the massless option and some recent conjectures on limiting behaviors that ought to allow ultraviolet completions~\cite{luest} although, as we have stressed, gravitino masses are expected to arise when radiative corrections are taken into account. This issue will be clearly worthy of further investigation once the complete spectrum and its stability properties will be addressed.} 

This discussion clearly needs to be complemented by a similarly detailed analysis of the bosonic spectrum, which will also highlight the implications for the perturbative stability of these vacua. This will be the subject of~\cite{ms22_2}. Moreover, a proper characterization of the effective four--dimensional theory, which appears to embody a non--linear realization of supersymmetry, cannot forego a similarly detailed analysis of the effects of radiative corrections, to which we hope to return in the near future.

\vskip 24pt
\section*{\sc Acknowledgments}

\vskip 12pt

We are grateful to G.~Dall'Agata and E.~Dudas for stimulating discussions, and also to S.~Raucci and Y.~Tatsuta, who read an earlier version of the manuscript and made useful comments. The work of AS was supported in part by Scuola Normale, by INFN (IS GSS-Pi) and by the MIUR-PRIN contract 2017CC72MK\_003. JM is grateful to Scuola Normale Superiore for the kind hospitality extended to him while this work was in progress. AS is grateful to Universit\'e de Paris Cit\'e and DESY--Hamburg for the kind hospitality, and to the Alexander von Humboldt foundation for the kind and generous support, while this work was in progress.
\newpage

\begin{appendices}

\section{\sc Fermi Couplings in the Einstein Frame} \label{app:fermiEinstein}
The fermionic equations that are relevant for our analysis can be conveniently extracted from~\cite{bergshoeff}, but we warn the reader that our $H$-terms have an overall factor of two with respect to their choice. In the string frame,  where the relevant contributions to the Lagrangian, obtained removing the two-forms and the axion, read~\footnote{In our conventions the antisymmetrization is such that $[AB]=AB - BA$.}
\begin{eqnarray}
{\cal L}_H &=&  \,-\,\frac{e}{2\,k_{10}^2}\ \Big\{e^{\,-\,2\,\phi} \Big[\ 2\,\bar{\psi}_M\,\Gamma^{MNP}\,D_N\,\psi_P \ - \ 2\,\bar{\lambda}\,\Gamma^M\,D_M\,\lambda \ + \ 4 \bar{\lambda}\,\Gamma^{MN}\,D_M\,\psi_N  \nonumber \\
&+& 4 \partial_M\,\phi \left(\bar{\psi}_N\,\Gamma^{N}\,\psi^M \ + \bar{\lambda} \,\Gamma^{N}\,\Gamma^{M}\,\psi_N\right) \Big] \nonumber \\ &+& e^{-\phi}\Big[\frac{1}{4}\, \bar{\psi}_M\,\Gamma^{[M} \, {\cal H}\!\!\!\!/ \ \Gamma^{B]}\,i\sigma_2\,\psi_B \,+\,
\frac{1}{2}\, \bar{\lambda}\,{\cal H}\!\!\!\!/ \ \Gamma^{B}\,i\sigma_2\,\psi_B \,-\, \frac{1}{4}\, \bar{\lambda}\,{\cal H}\!\!\!\!/ \ i\sigma_2\,\lambda\Big] \Big\} \ ,
\label{action_berg}
\end{eqnarray}
the relevant supersymmetry transformation rules of the Fermi fields are
\bea
\delta\,\psi_M &=& D_M\,\epsilon \ + \ \frac{e^\phi}{8}\, {\cal H}\!\!\!\!/ \ \Gamma_M\, i\,\sigma_2\, \epsilon \ , \nonumber \\
\delta\,\lambda &=& \Gamma^M\,\epsilon\,\partial_M\,\phi \  .
\eea
In this paper we are using a ``mostly plus'' signature, and the $\Gamma$'s are curved Dirac matrices. We shall use $\gamma$'s to denote flat Dirac matrices, when explicitly referring to the background. 

One can now reformulate these contributions in the Einstein frame as follows. To begin with, the vielbein and the covariant derivatives in the two frames are related according to
\beq
{e^{(s)}}_M^A = e^\frac{\phi}{4} \, e_M^A \ , \quad {D^{(s)}}_M \ = \ D_M \,+\, \frac{1}{8}\,\Gamma_{MN}\, \partial^N\,\phi \ .
\eeq
Moreover, it is convenient to redefine the Fermi fields as
\beq
\psi_M \ = \ e^{\frac{1}{8}\,\phi} \, {{\psi}'}_M \ , \quad \lambda \,=\, e^{-\frac{1}{8}\,\phi} \, \lambda' \ .
\eeq
so that the Lagrangian becomes
\begin{eqnarray}
{\cal L}_H &=&  \,-\,\frac{e}{2\,k_{10}^2}\ \Big\{ 2\,\bar{\psi}_M\,\Gamma^{MNP}\,D_N\,\psi_P \ - \ 2\,\bar{\lambda}\,\Gamma^M\,D_M\,\lambda \ + \ 4 \bar{\lambda}\,\Gamma^{MN}\,D_M\,\psi_N  \nonumber \\
&-& \frac{1}{2}\, \partial_M\,\phi \ \bar{\lambda} \,\Gamma^{N}\,\Gamma^{M}\,\psi_N \nonumber \\ &+& \frac{1}{4}\, \bar{\psi}_M\,\Gamma^{[M} \, {\cal H}\!\!\!\!/ \ \Gamma^{B]}\,i\sigma_2\,\psi_B \,+\,
\frac{1}{2}\, \bar{\lambda}\,{\cal H}\!\!\!\!/ \ \Gamma^{B}\,i\sigma_2\,\psi_B \,-\, \frac{1}{4}\, \bar{\lambda}\,{\cal H}\!\!\!\!/ \ i\sigma_2\,\lambda \Big\} \ ,
\label{action_berg2}
\end{eqnarray}
where for brevity we have omitted the ``primes'' on the new fields. The redefined supersymmetry transformations of the Fermi fields read
\bea
\delta\,{\psi}_M &=& D_M\,\epsilon \ + \ \frac{1}{8}\, {\cal H}\!\!\!\!/ \ \Gamma_M\, i\,\sigma_2\, \epsilon \ + \ \frac{1}{8}\,\Gamma^M\,\Gamma^N \,\epsilon\, \partial_N\,\phi  \ , \nonumber \\
\delta\,\lambda &=& \Gamma^M\,\epsilon\,\partial_M\,\phi \ , \nonumber \\
\eea
where the supersymmetry parameter was also redefined according to
\beq{}{}{}{}{}{}{}{}{}{}{}{}{}{}{}{}{}{}{}{}{}
\epsilon'=e^{-\frac{\phi}{8}}\,\epsilon \ .
\eeq
Now, however, we found it convenient to perform a further redefinition, introducing
\beq{}{}{}{}{}{}{}{}{}{}{}{}{}{}{}{}{}{}{}
{\psi''}_M \ = \ {\psi}_M \ - \ \frac{1}{8}\,\Gamma_M\,\lambda \ ,
\eeq
so that
\bea
\delta\,{\psi''}_M &=& D_M\,\epsilon \ + \ \frac{1}{8}\, {\cal H}\!\!\!\!/ \ \Gamma_M\, i\,\sigma_2\, \epsilon \ ,
\eea
and
\begin{eqnarray}
{\cal L}_H &=&  \,-\,\frac{e}{2\,k_{10}^2}\ \Big\{ 2\,\bar{\psi}_M\,\Gamma^{MNP}\,D_N\,{\psi}_P \ + \ \frac{1}{4}\,\bar{\lambda}\,\Gamma^M\,D_M\,\lambda  \nonumber \\
&-& \frac{1}{2}\, \partial_M\,\phi \ \bar{\lambda} \,\Gamma^{N}\,\Gamma^{M}\,{\psi}_N \,+\, \frac{1}{4}\, \bar{\psi}_M\,\Gamma^{[M} \, {\cal H}\!\!\!\!/ \ \Gamma^{B]}\,i\sigma_2\,{\psi}_B  \,+\, \frac{1}{16}\, \bar{\lambda}\,{\cal H}\!\!\!\!/ \ i\sigma_2\,\lambda \Big\} \ ,
\label{action_bergf}
\end{eqnarray}
where we have again removed the ``primes'' for brevity, while the supersymmetry transformations finally become
\bea{}{}{}{}{}{}{}{}{}{}{}{}{}{}{}{}{}{}
\delta\,{\psi}_M &=& D_M\,\epsilon \ + \ \frac{1}{8}\, {\cal H}\!\!\!\!/ \ \Gamma_M\, i\,\sigma_2\, \epsilon \ , \nonumber \\
\delta\,\lambda &=& \Gamma^M\,\epsilon\,\partial_M\,\phi \ . \label{susy_final}
\eea
These results would be of direct relevance if $\phi$ were not constant.
The fermionic equations of motion that we refer to in the main body of the paper thus read
\bea{}{}{}{}{}{}{}{}{}{}{}{}{}{}{}{}{}{}
&& {\Gamma^{MNP}\,D_N\,\psi_P \ + \ \frac{1}{8}\,\Gamma^{[M}\,{\cal H}\!\!\!\!/ \ \Gamma^{N]}\,i\,\sigma_2\,\psi_N}  \ - \ \frac{1}{8}\,\partial_N\,\phi\,\Gamma^N\,\Gamma^M\lambda   \ = \ 0 \ , \label{grav_eq_final}   \nonumber \\
&& \Gamma^M\,D_M\,\lambda \ - \ \partial_M\phi\,\Gamma^{N}\,\Gamma^M\psi_N \ + \ \frac{1}{4}\,{\cal H}\!\!\!\!/ \ i\,\sigma_2\,\lambda \ = \ 0 \ . \label{eqs_fermi_app}
\eea

In our background
\beq
{\cal H}\!\!\!\!/ \ \equiv \ \frac{1}{5!}\,{\cal H}_{5\,M_1\cdot M_5}\,\Gamma^{M_1 \cdots M_5} \ = \ - \ 2\,H \,e^{-5C}\,i\sigma_2\,\Lambda\,\gamma_r \left( \frac{1 \,-\, \gamma_{11}}{2} \right) \label{Hslash}
\eeq
when acting on a chiral ten--dimensional spinor.
Moreover
\bea
&& \frac{1}{4}\,{\omega_\mu}^{AB}\,\gamma_{AB} \ = \ \frac{1}{2}\ \gamma_\mu \gamma_r\ e^{A-B}\, A' \ = \  \frac{1}{2}\ \gamma_\mu \gamma_r\ e^{-3 A-5 C}\, A' \ , \nonumber \\
&& \frac{1}{4}\,{\omega_{i}}^{AB}\,\gamma_{AB} \ = \  \frac{1}{2}\ \gamma_i \gamma_r\ e^{C-B}\, C' \ = \ \frac{1}{2}\ \gamma_i \gamma_r\ e^{-4 A-4 C}\, C' \ , \label{omegaab}
\eea
while
\beq
\Lambda \ = \ \gamma^{0123}\,i\,\sigma_2 \ , \label{Lambda}
\eeq
and
\beq{}{}{}{}{}{}{}{}{}{}{}{}{}{}{}{}{}{}{}{}
B \ = \ 4 \,A \ + \ 5\, C \ ,
\eeq
in the harmonic gauge.

In detail, taking into account that, as explained in Section~\ref{sec:background}, in our backgrounds the dilaton profile is constant, and using the $z$-variable introduced in eq.~\eqref{dzdr} and the definition of ${\cal W}_5$ given in eq.~\eqref{calH5},
\bea
&& \delta\,\psi_\mu  \ = \ \partial_\mu\,\epsilon \ + \ \frac{1}{2}\,\gamma_\mu\,\gamma_r\,\Big(A_z \ + \ {\cal W}_5\,\Lambda \Big)\, \epsilon \ , \nonumber \\
&& \delta\,\psi_r \ = \ e^{B-A} \left(\partial_z \ + \ \frac{1}{2}\, {\cal W}_5\,\Lambda\right)\epsilon\ , \nonumber \\
&& \delta\,\psi_i \ = \ \frac{1}{2}\,\gamma_i\,\gamma_r e^{C-A}\,\Big(C_z \ - \ {\cal W}_5\,\Lambda \Big)\, \epsilon\ .
\eea

\subsection{\sc Spin--$\frac{3}{2}$ Equations} \label{app:fermiEinstein32}

In this case we let $\psi_r=\psi_i=0$ and work with only $\psi_\mu$. Consequently the gravitino kinetic terms yield the contributions
\bea
\Gamma^{\mu NP}\,D_N\,\psi_P &=& e^{-3A}\left[ \gamma^{\mu\nu\rho}\,\partial_\nu \psi_\rho  \ - \  \gamma^{\mu\rho}\,\gamma_r \left( \partial_z\,+\,A_z+ \frac{5}{2}\,C_z\right)\psi_\rho \right]
\ , \nonumber \\
\Gamma^{r NP}\,D_N\,\psi_P &=& \, e^{-2A-B}\left[\gamma^r \,\gamma^{\mu\rho} \ \partial_\mu\,\psi_\rho \ + \ \left(\frac{3}{2}\,A_z+ \frac{5}{2}\,C_z\right)\gamma^\rho\,\psi_\rho\right] \label{app:kin32} \\
\Gamma^{i NP}\,D_N\,\psi_P &=&  e^{-C-2A}\left[\gamma^i\,\gamma^{\mu\rho} \ \partial_\mu\,\psi_\rho \ + \ \gamma^i\,\gamma_r\, \left(\partial_z \,+\,  \frac{3}{2}\,A_z\,+\,2\,\,C_z\right) \, \gamma^\rho\,\psi_\rho \right]
\ , \nonumber
\eea
which are used in Section~\ref{sec:Fermi modes}.

Similarly, the ${\cal H}\!\!\!\!/$ terms entering the Rarita--Schwinger equation are
\bea{}
\frac{1}{8\,}\,\Gamma^{[M}\,{\cal H}\!\!\!\!/ \ \Gamma^{N]}\,i\,\sigma_2\,\psi_N \ = \   \frac{H}{4}\,e^{-5C}\left(\Gamma^{M}\,\Lambda\,\gamma_r\, \Gamma^{N} \ - \ \Gamma^{N}\,\Lambda\,\gamma_r\,\Gamma^{M}\right) \,\psi_N \ ,
\eea
and the three distinct index ranges for $M$ give
\bea{}
&&\frac{1}{8\,}\,\Gamma^{[\mu}\,{\cal H}\!\!\!\!/ \ \Gamma^{N]}\,i\,\sigma_2\,\psi_N \ = \  \frac{H}{2}\,e^{-2A-5C} \,\gamma_r\,\gamma^{\mu\nu}\,\Lambda\,\psi_\nu  \ , \nonumber \\
&&\frac{1}{8\,}\,\Gamma^{[r}\,{\cal H}\!\!\!\!/ \ \Gamma^{N]}\,i\,\sigma_2\,\psi_N \nonumber\ = \ \frac{H}{2}\,e^{-A-B-5C} \,\gamma^{\nu}\,\Lambda\,\psi_\nu \ , \nonumber \\
&&\frac{1}{8\,}\,\Gamma^{[i}\,{\cal H}\!\!\!\!/ \ \Gamma^{N]}\,i\,\sigma_2\,\psi_N \ = 0 \ . \label{app:H32}
\eea

The three components of the gravitino equation are thus
\bea
&& \gamma^{\mu\nu\rho}\,\partial_\nu \psi_\rho  \, - \, \gamma^{\mu\rho}\,\gamma_r \left( \partial_z\,+\,A_z+ \frac{5}{2}\,C_z\right)\psi_\rho \,+\, {\cal W}_5\,\gamma_r\,\gamma^{\mu\nu}\,\Lambda\,\psi_\nu \ = \ 0 \ , \nonumber \\
&& \gamma^r \,\gamma^{\mu\rho} \ \partial_\mu\,\psi_\rho \, + \, \left(\frac{3}{2}\,A_z+ \frac{5}{2}\,C_z\right)\gamma^\rho\,\psi_\rho \,+\, {\cal W}_5 \,\gamma^{\nu}\,\Lambda\,\psi_\nu \ , \nonumber \\
&& \gamma^i\,\gamma^{\mu\rho} \ \partial_\mu\,\psi_\rho \ + \ \gamma^i\,\gamma_r \left(\partial_z \,+\,  \frac{3}{2}\,A_z\,+\,2\,\,C_z\right) \, \gamma^\rho\,\psi_\rho \ = \ 0 \ . \label{eqs32}
\eea

\subsection{\sc Spin--$\frac{1}{2}$ Modes in the Spinorial of $SO(6)$}\label{app:fermiEinstein12}

To begin with, after a gauge choice one is left with four types of these spin--$\frac{1}{2}$ modes:
\bea
&& \psi_\mu \ = \  \gamma_\mu \chi_1  \ , \qquad \psi_r = \gamma_r\,{\chi}_2  \ , \qquad \psi_i \ = \ \gamma_i\,{\chi}_3 \ , \qquad  \lambda \ ,
\eea

The gravitino kinetic term now gives
\bea
 \Gamma^{MNP}\,D_N\,\psi_P &=& \Gamma^{M \nu \rho}\left(\partial_\nu + \frac{1}{2}\gamma_\nu\,\gamma_r\,e^{A-B}\,A'\right)\gamma_\rho\,\chi_1 \nonumber \\ &+&
\Gamma^{M \nu r}\left(\partial_\nu + \frac{1}{2}\gamma_\nu\,\gamma_r\,e^{A-B}\,A'\right)\gamma_r\, \chi_2  \\
&+& \Gamma^{M \nu i}\left(\partial_\nu + \frac{1}{2}\gamma_\nu\,\gamma_r\,e^{A-B}\,A'\right)\gamma_i\,\chi_3\,+\,  \ \Gamma^{M r \rho}\,\gamma_\rho\,\partial_r\,\chi_1  \ + \ \Gamma^{M r i}\,\partial_r\,\gamma_i\,\chi_3  \nonumber \\
&+& \frac{1}{2}\,e^{C-B}\,C'\Big[\Gamma^{M i \rho}\,\gamma_i\,\gamma_r\,\gamma_\rho\,\chi_1  \,+\, \Gamma^{M i r}\,\gamma_i\,\gamma_r\,\gamma_r\, \chi_2 \,+\, \Gamma^{M i j}\,\gamma_i\,\gamma_r\gamma_j\,\chi_3\Big] \ , \nonumber
\eea
and in detail
\bea
 \Gamma^{\mu NP}\,D_N\,\psi_P &=& e^{-2A}\gamma^{\mu\nu}\,\partial_\nu \left( 2 e^{-A}\,\chi_1 + e^{-B}\,\chi_2 +5 e^{-C}\,\chi_3\right) \nonumber \\
&+& \frac{3}{2}\,e^{-A-B}\,A'\,\gamma^\mu\,\gamma_r \left( 2 e^{-A}\,\chi_1 - e^{-B}\,\chi_2 + 5\,e^{-C}\,\chi_3 \right) \nonumber \\ &+& e^{-A-B}\left( 3\,e^{-A}\,\gamma^{\mu r}\,\partial_r\,\chi_1 + 5\,e^{-C}\,\gamma^{\mu r}\,\partial_r\,\chi_3\right) \nonumber \\
&+& \frac{5}{2}\,e^{-A-B}\,C'\,\gamma^\mu\,\gamma_r\left( 3\, e^{-A}\,\chi_1\,-\,e^{-B}\,\chi_2 \,+\, 4\, e^{-C}\,\chi_3\right)  \ ,
\label{spacetime_RS_spinor}
\eea
\bea
 \Gamma^{r NP}\,D_N\,\psi_P &=& e^{-A-B}\left( \gamma^{r\nu}\,\partial_\nu + 2 e^{A-B}\,A'\right)\left( 3 e^{-A}\,\chi_1+ 5 e^{-C}\chi_3\right) \nonumber \\
&+& 10\,e^{-2B}\,C'\left( e^{-A}\chi_1\,+\, e^{-C}\chi_3\right) \ ,
\label{r_RS_spinor}
\eea
\bea
 \Gamma^{k NP}\,D_N\,\psi_P &=& e^{-A-C}\,\gamma^{k\nu}\,\partial_\nu \left( 3\,e^{-A}\,\chi_1 +e^{-B}\,\chi_2 + { 4} \, e^{-C}\,\chi_3\right)  \nonumber \\
&+& 2\,e^{-B-C} A'\gamma^k\,\gamma_r \left( 3\,e^{-A}\,\chi_1 -e^{-B}\,\chi_2 + { 4} \, e^{-C}\,\chi_3\right)  \nonumber \\
&+& e^{-B-C} \left(4 e^{-A} \gamma^{k r}\,\partial_r \,\chi_1 + {4} \, e^{-C}\gamma^{kr}\,\partial_r\,\chi_3 \right) \nonumber \\
&+&  2\,e^{-B-C}\,C'\, \gamma^k\gamma_r\left(4\,  e^{-A} \chi_1\,-e^{-B} \chi_2 \,+3 \, e^{-C}\chi_3\right)  \ . \label{internal_RS_spinor}
 \eea
 Moreover
 \bea{}{}{}{}{}{}{}{}{}{}{}{}{}{}{}{}
 \Gamma^M\,D_M\,\lambda \ = \ e^{-A}\gamma^\mu\,\partial_\mu\,\lambda \ + \ e^{-B} \gamma^r\left(\partial_r \, + \, 2 A'\,+\, \frac{5}{2}\,C'\right)\lambda \ ,
 \eea
and expanding the form couplings now gives
\bea{}
&&\frac{1}{8\,}\,\Gamma^{[\mu}\,{\cal H}\!\!\!\!/ \ \Gamma^{N]}\,i\,\sigma_2\,\psi_N \ = \  \frac{H}{4}\,e^{-5C}\left(\Gamma^{\mu}\,\Lambda\,\gamma_r\, \Gamma^{N} \ - \ \Gamma^{N}\,\Lambda\,\gamma_r\, \Gamma^{\mu}\right) \,\psi_N \nonumber \\
&&=\ \frac{H}{2}\,e^{-2A-5C} \,\gamma_r\,\gamma^{\mu\nu}\,\Lambda\,\psi_\nu \ + \ \frac{H}{2}\,e^{-A-B-5C} \,\gamma^{\mu}\,\Lambda\,\psi_r \ , \nonumber \\
&&\frac{1}{8\,}\,\Gamma^{[r}\,{\cal H}\!\!\!\!/ \ \Gamma^{N]}\,i\,\sigma_2\,\psi_N \nonumber\ = \ {}\, \frac{H}{2}\,e^{-A-B-5C} \,\gamma^{\nu}\,\Lambda\,\psi_\nu \ , \nonumber \\
&&\frac{1}{8\,}\,\Gamma^{[i}\,{\cal H}\!\!\!\!/ \ \Gamma^{N]}\,i\,\sigma_2\,\psi_N \ = \ - \ \frac{H}{2}\,e^{-7C} \,\gamma_r\,\gamma^{ij}\,\Lambda\,\psi_j \ ,
\eea
or in terms of the $\chi_i$
\bea{}
&&\frac{1}{8\,}\,\Gamma^{[\mu}\,{\cal H}\!\!\!\!/ \ \Gamma^{N]}\,i\,\sigma_2\,\psi_N \ = \ \frac{H}{2}\,e^{-A-5C} \,\gamma^{\mu}\,\gamma_r\,\Lambda\left( 3\,e^{-A} \,\chi_1 \ { +} \ e^{-B} \,\chi_2\right) \ , \nonumber \\
&&\frac{1}{8\,}\,\Gamma^{[r}\,{\cal H}\!\!\!\!/ \ \Gamma^{N]}\,i\,\sigma_2\,\psi_N \nonumber\ = \ 2\,H\,e^{-A-B-5C} \,\Lambda\,\chi_1 \ , \nonumber \\
&&\frac{1}{8\,}\,\Gamma^{[i}\,{\cal H}\!\!\!\!/ \ \Gamma^{N]}\,i\,\sigma_2\,\psi_N \ =  2\,H\,e^{-7C} \,\gamma^{i}\,\gamma_r\,\Lambda\,\chi_3 \ , \nonumber \\
&& \frac{1}{4}\,{\cal H}\!\!\!\!/ \  i \, \sigma_2\,\lambda \ = \ \frac{H}{2}\,e^{ - 5 C}\,\gamma_r\,\Lambda\,\lambda \ .
\eea

Consequently, in the backgrounds of interest the dilatino equation for spin--$\frac{1}{2}$ modes is
\bea
&& \gamma^\mu\,\partial_\mu\,\lambda \ + \ \gamma^r\left(\partial_z \, + \, 2 A_z\,+\, \frac{5}{2}\,C_z\right)\lambda \ + \ {\cal W}_5\,\gamma_r\,\Lambda\,\lambda \nonumber \\
&+& \phi_z\,\gamma_r\left(4\,e^{-A}\,\chi_1\,-\,e^{-B}\,\chi_2\,+\,5\,e^{-C}\,\chi_3 \right) \ = \ 0 \ .
\eea

In a similar fashion, the complete equations originating from the gravitino are
\bea{}{}{}
&& \gamma^{\mu\nu}\,\partial_\nu \left( 2 e^{-A}\,\chi_1 + e^{-B}\,\chi_2 +5 e^{-C}\,\chi_3\right) \,+\, \frac{3}{2}\,A_z\,\gamma^\mu\,\gamma_r \left( 4 e^{-A}\,\chi_1 - e^{-B}\,\chi_2 + 5\,e^{-C}\,\chi_3 \right) \nonumber \\ &+& \gamma^\mu\,\gamma_r\,\partial_z\left( 3\, e^{-A}\chi_1 + 5\,e^{-C}\chi_3\right) \,+\, \frac{5}{2}\,C_z\,\gamma^\mu\,\gamma_r\left( 3\, e^{-A}\,\chi_1\,-\,e^{-B}\,\chi_2 \,+\, 6\, e^{-C}\,\chi_3\right) \nonumber \\ &{+}& {\cal W}_5 \,\gamma^{\mu}\,\gamma_r\,\Lambda\left( 3\,e^{-A} \,\chi_1 \ {+} \ e^{-B} \,\chi_2\right) \,+\, \phi_z\,\gamma^\mu\,\gamma_r\,\lambda \ = \ 0 \ , \label{grav12mu}
\eea
\bea
&& \left( \gamma^{r\nu}\,\partial_\nu + 2 e^{A-B}\,A'\right)\left( 3 e^{-A}\,\chi_1+ 5 e^{-C}\chi_3\right) \nonumber \\
&+& 10\,C_z\left( e^{-A}\chi_1\,+\, e^{-C}\chi_3\right) \ + \  4\,{\cal W}_5 \,\Lambda\,e^{-A}\,\chi_1  \,-\, \frac{1}{8}\ \,\phi_z\,\lambda\ = \ 0 \ ,  \label{grav12r}
\eea
\bea
&& \gamma^{k\nu}\,\partial_\nu \left( 3\,e^{-A}\,\chi_1 +e^{-B}\,\chi_2 + { 4} \, e^{-C}\,\chi_3\right)  \,+\, 2\,A_z\gamma^k\,\gamma_r \left( 5\,e^{-A}\,\chi_1 -e^{-B}\,\chi_2 + { 4} \, e^{-C}\,\chi_3\right)  \nonumber \\
&+& 4\,\gamma^k\gamma_r\,\partial_z\left(e^{-A}\chi_1\,+ \, e^{-C}\chi_3\right) \,+\,  2\,C_z\, \gamma^k\gamma_r\left( 4\,  e^{-A} \chi_1\,-e^{-B} \chi_2 \,+5 \, e^{-C}\chi_3\right)  \nonumber \\
&+& 4 \,{\cal W}_5 \,e^{-C}\,\gamma^{k}\,\gamma_r\,\Lambda\,\chi_3\,+\,\frac{1}{8}\ \phi_z\,\gamma^k\,\gamma_r\,\lambda \ = \ 0 \ . \label{grav12i}
\eea
\subsection{\sc Spin--$\frac{1}{2}$ Modes from $\psi_i$} \label{app:fermimodes12i}

We can finally consider the contribution of $\psi_i$, with $\gamma^i\,\psi_i = 0$. The starting point is provided by
\bea
&& \Gamma^{\mu N i }\,D_N\,\psi_i  \ = \ 0 \ , \nonumber \\
&& \Gamma^{r N i }\,D_N\,\psi_i \ = \ 0 \ , \nonumber \\
&& \Gamma^{k N i }\,D_N\,\psi_i  \ + \ \frac{1}{8}\,\Gamma^{[k}\,{\cal H}\!\!\!\!/ \ \Gamma^{i]}\,i\,\sigma_2\,\psi_i \ = \ 0 \ ,
\eea
since, as we have seen, in the first two cases the term proportional to $H$ vanishes. The first two equations are identically satisfied, while the last becomes
\beq{}{}{}{}{}{}{}{}{}{}{}{}{}{}{}{}{}{}{}{}{}
\left[\partial \!\!\! / \, + \, \gamma_r\left(\partial_z\,+\,2\,A_z \,+\, \frac{3}{2}\,C_z\,+\, {\cal W}_5 \,\Lambda \right)\right]\,\psi^k  \ = \ 0 \ .
\eeq

\section{\sc Fermi Modes in an interval} \label{app:fermi_interval}

In this paper we were confronted with metrics of the form
\beq
ds^2 \ = \ e^{2\,A(r)}\, dx^2 \ + \ e^{2\,B(r)}\, dr^2 \ + \ e^{2\,C(r)}\, dy^2 \ , \label{metric_ABC}
\eeq
which describe in the $r$ direction segments of finite length, or at least half--lines. In all these cases the Dirac or Rarita--Schwinger equations must be supplemented by boundary conditions, as described in~\cite{boundary}. Here we would like to elaborate on their implications, focusing on the simplest case, the Dirac equation for the isotropic metric (with no $y$--coordinates) in conformal gauge, which reads
\beq
ds^2 \ = \ e^{2\,\Omega(z)}\left( dx^2 \ + \  dz^2 \right) \ .
\eeq
To begin with, let us notice that the spin connection one--form for this class of metrics has the non--vanishing components
\beq
{\omega}^{\nu z} \ = \ dx^\mu\, \delta_\mu^\nu\, \Omega'(z) \ ,
\eeq
so that the massless Dirac action reads~\footnote{Here we are reverting to the more conventional choice of introducing an overall factor of i in front of the action. The difference with the conventions drawn from~\cite{bergshoeff} and used in Appendix~\ref{app:fermiEinstein} resides in a different definition of conjugation for Fermi fields: here $(AB)^\dagger = B^\dagger\,A^\dagger$, or alternatively in the introduction of a factor $i$ in the definition of $\bar{\psi}$.}
\beq
{\cal S} \ = \ - \ i\int d^{D-1}x \ dz \ e^{(D-1)\Omega}\ \bar{\psi} \left[ \gamma^z\left( \partial_z \ + \frac{D-1}{2}\ \Omega'  \right) \ + \ \gamma^\mu\, \partial_\mu\right] \psi \ ,
\eeq
where all $\gamma$--matrices have flat indices. The redefinition
\beq
\psi \ = \ e^{\,-\,\frac{(D-1)}{2}\,\Omega} \, \chi
\eeq
leads therefore to a free Fermi problem, with
\beq
{\cal S} \ = \ - \ i\int d^{D-1}x \,dz \ \bar{\chi} \big[ \gamma^z\,\partial_z  \ + \ \gamma^\mu\, \partial_\mu\big] \chi \ ,
\eeq
and the Dirac equation for $\chi$ thus becomes
\beq
\big[ \gamma^z\,\partial_z  \ + \ \gamma^\mu\, \partial_\mu\big] \chi(x,z) \ = \ 0 \  .
\eeq
Since the explicit $z$ dependence has disappeared, it is convenient to focus on eigenstates of $\,-\,i\partial_z$ of eigenvalue $k$, which are of the form
\beq
\chi(x,z) \ = \ \chi_k(x) \, e^{i k z} \ . \label{modes}
\eeq
For them, the Dirac equation reduces to
\beq
\gamma^\mu\, \partial_\mu \, \chi_k(x) \ = \ - i\, k \, \gamma^z\,\chi_k(x) \ , \label{dirac_k}
\eeq
and consequently
\beq
\Box \, \chi_k(x) = k^2 \,\chi_k(x) \ ,
\eeq
so that $\chi_k(x)$ and $\chi_{-k}(x)$ correspond to the same mass $m=\left|k\right|$.

In the presence of only one internal dimension associated to $z$, one can distinguish four cases.
\begin{enumerate}
    \item \emph{An infinite line}. In this case $k$ is an arbitrary real number and there is a continuous mass spectrum. The absence of a mass gap reflects the $D$-dimensional nature of the system.
    \item \emph{A circle}. In this case $z$ is identified with $z+2\pi R$, and there are further distinctions.  To begin with, for a Dirac spinor one can demand, in general, that
    \beq \chi(x,z+2\pi R) \ = \ e^{i\,2\,\pi\,\alpha}\,\chi(x,z)  \ , \label{quasi_periodicity} \eeq
    with an arbitrary value of $\alpha$, consistently with the periodicity of all fermionic bilinears. This condition determines
    $$ k \ = \ \frac{n\ + \ \alpha}{R}\qquad n \in Z \ , $$
    with an arbitrary integer value of $n$. For a Majorana spinor, whenever it can be defined, pairs of modes of this type, with opposite values of $k$, must combine to grant that
    \beq \chi(x,z) \ = \ C \, \overline{\chi}^T(x,z) \label{majoranaeq} \ . \eeq
    This condition, however, can be consistently imposed only if $\alpha=0,\frac{1}{2}$, since only in these cases modes with opposite values of $k$ exist. One can then combine pairs of modes of mass $m$ according to
    \beq \chi(x,z) \ = \ \chi_m(x) \,e^{i\,m\,z} \ + \ \chi_{-m}(x)\,e^{\,-\,i\,m\,z} \ , \label{combin} \eeq

    and eq.~\eqref{majoranaeq} can be satisfied demanding that
    \beq \chi_{-m}(x) \ = \ C \,\overline{\chi}_m^T(x) \ . \label{majorana} \eeq

    \item \emph{A semi--infinite line}. We set the boundary at $z=0$, where one is to enforce the boundary condition
\beq
\chi(x,0)\ = \ \gamma^z \, \chi(x,0) \ . \label{bcz0}
\eeq
Eq.~\eqref{dirac_k}, however, implies that $\gamma^z\,\chi_k(x)$ satisfies the same equation as $\chi_{-k}(x)$, since
\beq
\gamma^\mu\, \partial_\mu \, \left(\gamma^z\,\chi_k(x)\right) \ = \ i\, k \, \gamma^z\left(\gamma^z\,\chi_k(x)\right) \ , \label{dirac_kz}
\eeq
so that one cannot demand that $\chi_k(x)$ be an eigenstate of $\gamma^z$, as in the boundary condition~\eqref{bcz0}, if $k \neq 0$. Rather, a given mode of mass $m$ must be a combination of a pair of these functions, as in eq.~\eqref{combin}. It is thus convenient to define
    \beq \chi_m^\pm(x) \ = \ \chi_m(x) \ \pm \ \chi_{-m}(x) \ ,\eeq
    and then the boundary condition~\eqref{bcz0} translates into
    \beq
    \gamma^z\, \chi_m^+ \ = \ \chi_m^+\ .  \label{gammachiplus}
    \eeq
    On the other hand, the Dirac equation can be recast in the form
    \beq
    \gamma^\mu\,\partial_\mu\, \chi_m^\pm \ = \ - \ i\,m\,\gamma^z \, \chi_m^\mp \ , \label{diracpm}
    \eeq
   and together with eq.~\eqref{gammachiplus} implies that
     \beq\gamma^z\, \chi_m^- \ = \ - \ \chi_m^- \ .
    \eeq
   As a result, the Dirac equation for these combinations takes finally the simpler form
    \beq
    \gamma^\mu\,\partial_\mu\, \chi_m^\pm \ = \ \pm \ i\,m \, \chi_m^\mp \ ,
    \eeq
    while the independent modes are conveniently expressed as
    \beq
    \chi_m(x,z) \ = \ \chi_m^+(x) \,\cos mz \ + \ i \,  \chi_m^-(x) \,\sin mz \ . \label{oneboundary}
    \eeq
   This would be a possible choice even in the absence of a boundary, of course, and would go along with the corresponding form~\eqref{diracpm} of the Dirac equation. The boundary, however, introduces the two projections
    \beq\gamma^z\, \chi_m^\pm \ = \ \pm \ \chi_m^\pm  \label{projections}
    \eeq
    on $\chi^\pm$, at $z=0$. Moreover, when the Majorana condition~\eqref{majorana} holds,  $\chi_m^+$ is real while $\chi_m^-$ is purely imaginary in a Majorana representation, and $\chi_m(x,z)$ is then real.

\item \emph{A finite $z$-interval}, $0 \leq z \leq z_m$. As in the previous case, the boundary condition at $z=0$ leads to eqs.~\eqref{oneboundary} and \eqref{projections}, but one must demand, in addition, that
\beq
\gamma^z \, \chi_m(x,z_m)  \ = \ \pm \ \chi_m(x,z_m)  \ .
\eeq
According to eq.~\eqref{projections}, the upper sign requires that
\beq
\sin \left(m \,z_m\right) \ = \ 0 \ , \qquad m \ =\ \frac{n\,\pi}{z_m} \ , \qquad n=1,2,\ldots\ ,
\eeq
and leads to the Ramond sector. On the other hand, the lower sign choice requires that
\beq
\cos\left( m\,z_m\right) \ = \ 0 \ , \qquad m \ =\ \frac{n+\frac{\pi}{2}}{z_m} \ , \qquad n=0,2,\ldots\ ,
\eeq
and leads to the Neveu--Schwarz sector. In the main body of the paper we focused on the first option, which led to a residual massless spectrum of Fermi fields. The other choice, which would have given mass to these fields, would have implemented the Scherk--Schwarz mechanism on the low--energy effective field theory.
\end{enumerate}

Notice that, in  the metrics in eq.~\eqref{metric_ABC}, one can compute $z_m$ as
\beq
z_m \ = \ \int \, dr\ e^{B-A} \ .
\eeq
However, when $A \neq 0$, as in the main body of the paper, $z_m$ is not the actual length of the interval, which is determined by
\beq
\int \, dr\ e^{B} \ .
\eeq
If $z_m$ is finite, we are in case 4, while if $z_m$ is infinite, we are in case 3.

As we stressed in~\cite{boundary}, in many cases of interest for String Theory, eq.~\eqref{bcz0} does not admit solutions compatible with maximal symmetry in the $x$ directions, due to chirality and/or Majorana constraints. However, if the interval is combined with an internal torus, solutions of the constraints of the type
\beq
\chi(x,0) \ = \ \Lambda \, \chi(x,0) \ ,
\eeq
can exist, with a Hermitian matrix $\Lambda$ such that
\beq
\Lambda^2 \ = \ 1\ , \qquad \left\{ \Lambda \,,\, \gamma^0\gamma^r\right\} \ = \ 0 \ , \label{Lambda_constr_kin}
\eeq
which are compatible with lower--dimensional Lorentz symmetries. The preceding considerations apply to these more complicated cases, provided one includes in $m$ the contributions of toroidal modes and $\gamma^z$ is replaced by $\Lambda$.

In type IIB compactified on a five--torus $T^5$, where the Fermi fields are doublets of Majorana--Weyl spinors, as we have seen
\beq
\Lambda \ = \ \gamma^0\,\ \gamma^1 \,\ \gamma^2\,\ \gamma^3\,i\,\sigma^2 \ ,
\eeq
which is Hermitian, Lorentz invariant and satisfies the constraint in eq.~\eqref{Lambda_constr_kin}. Notice that this choice is slightly more general than those considered in~\cite{boundary}, where the main focus was on orientifolds, since it mixes the pairs of original spinor fields of the ten--dimensional IIB supergravity.

\section{\sc A Recurrent Sturm--Liouville Problem} \label{app:sturmliouville}

In this Appendix we summarize the properties of a Sturm--Liouville problem that surfaces, in different forms, in our analysis. The relevant structure came to the forefront, in Physics, in Witten's work on the dynamical breaking of supersymmetry (see~\cite{witten_dyn} and references therein), and our main target here is the role of boundary conditions, following~\cite{boundary}, since in our case the actual spectra emerging from these problems depend crucially on them. We also distinguish the two presentations of the problem that show up with Fermi and Bose fields, which are related but not equivalent. The latter type of fields will be the main subject of the companion papers~\cite{ms22_2}.

\subsection{\sc Fermi Fields}

In this paper we often deal with a system of differential equations of the form
\bea
&& {\cal A} \,\psi_{1\,m} \ = \ m\, \psi_{2\,m} \ , \nonumber \\
&& {\cal A}^\dagger\,\psi_{2\,m}  \ = \ m \, \psi_{1\,m}  \ , \label{SL_sys}
\eea
which emerge from Fermi fields in the presence of an internal interval after separation of variables,
where
\bea{}{}{}
{\cal A} &=& \partial_z \ + \ {\cal W}_z \ , \nonumber \\
{\cal A}^\dagger &=& - \ \partial_z\ + \ {\cal W}_z \ .
\eea
The system can be presented in the manifestly Hermitian matrix structure
\beq{}{}{}
{\cal Q} \, \Psi_m \ = \ m\, \Psi_m \ ,
\eeq
with
\beq{}{}{}
{\cal Q}\,=\,\left(\begin{array}{cc} 0 & {\cal A}^\dagger \\ {\cal A} & 0 \end{array}\right) \ , \qquad \Psi_m \,=\, \left(\begin{array}{c} \psi_{1m} \\ \psi_{2m} \end{array} \right) \ .
\eeq
Clearly, if ${\cal Q}$ is Hermitian $m^2 \geq 0$. This condition demands that
\beq{}{}{}{}
\left(\Psi_m,{\cal Q}\,\Psi_{m'}\right) \ = \ \left({\cal Q}\,\Psi_m,\Psi_{m'}\right) \ , \label{Qmm'}
\eeq
for all choices of $m$ and $m'$, with the standard scalar product
\beq{}{}{}{}{}
\left(\Psi,\Psi'\right) \ = \  \int_0^{z_m} dz \ \Psi^\dagger \, \Psi' \ .
\eeq
Hermiticity thus holds provided the boundary conditions
\beq{}{}{}{}
\left[ \Psi_{m}^\dagger \, \sigma_2\,\Psi_{m'}\right]_0^{z_m}  \ = \ i \Big[  \psi_{2m}^\star\,\psi_{1m'} \ - \ \psi_{1m}^\star\,\psi_{2m'} \Big]_0^{z_m}  \ = \ 0  \label{bc_sigma2}
\eeq
hold at the two ends of the interval. In particular, the choice
\beq{}{}{}{}{}
\psi_{2m} \ = \ 0  \label{psi20_bc}
\eeq
at the boundaries clearly solves eq.~\eqref{bc_sigma2}. In the main body of the paper, this choice resulted from working with eigenstates of $\Lambda$, compatibly with the no--flow conditions of~\cite{boundary}. More generally, one could demand that, at the boundary,
\beq{}{}{}{}
\Psi_m \ = \ \left(\sin\theta \,\sigma_1 \ + \ \cos\theta \, \sigma_3\right)\Psi_m  \ ,
\eeq
with $\sigma_1$ and $\sigma_3$ Pauli matrices $\theta$ an arbitrary angle, while eq.~\eqref{psi20_bc} corresponds to $\theta=0$. When this type of condition holds, eq.~\eqref{Qmm'} implies that
\beq{}{}{}{}
\left(\Psi_m,\Psi_{m'}\right) \ = \ 0   \label{Fermi_bc}
\eeq
for $m \neq m'$.
\subsection{\sc Bose Fields}

The analysis of Bose fields in the companion papers~\cite{ms22_2}, which already surfaced in Section~\ref{sec:tricky_zero_mode}, leads to Schr\"odinger--like equations of the form
\beq{}{}{}{}
{\cal A}^\dagger\, {\cal A}\,\chi \ = \ m^2\, \chi \ , \label{second_order}
\eeq
for a single--component field $\chi$, and to generalizations of this problem to the multi--field case.  For $m \neq 0$, the system~\eqref{second_order} is equivalent to studying the fermionic system~\eqref{SL_sys}, with
\beq{}{}{}{}{}
\psi_1 \ = \ \frac{1}{\sqrt{2}}\,\chi\, \qquad \psi_2 \ = \ \frac{1}{\sqrt{2}\,m}\, {\cal A}\,\chi \ .
\eeq
and the boundary condition~\eqref{psi20_bc} translates into the demand that
\beq{}{}{}{}
{\cal A}\,\chi \ = \ 0  \label{bc_chi}
\eeq
at the boundary.
However, when $m=0$ the Fermi system~\eqref{SL_sys} decouples in the two equations
\bea
{\cal A}\,\psi_{1\,0} &=& 0 \label{gp_sys0}\ , \nonumber \\
{\cal A}^\dagger\,\psi_{2\,0} &=&  0  \ , \label{SL_sys0}
\eea
which lead to the two types of ground--state wavefunctions,
\beq{}{}{}{}
\psi_{10} \ = \ c_1\ e^{-\, {\cal W}} \ , \qquad \psi_{20} \ = \ c_2\ e^{ \,{\cal W}} \ .
\eeq
Their normalizability depends on the choice of ${\cal W}$, and the Fermi boundary conditions~\eqref{Fermi_bc} with $\theta=0$ would remove $\psi_{20}$.

Summarizing, the Fermi and Bose problems are in one-to-one correspondence for nonzero values of $m$, provided the two boundary conditions~\eqref{psi20_bc} and~\eqref{bc_chi} are enforced. On the other hand, if~\eqref{second_order} is the starting point,
\beq{}{}{}{}{}
\chi \ = \ \psi_{10}
\eeq
is the actual ground state if it is normalizable, while the boundary condition~\eqref{bc_chi} is identically satisfied. When the Schr\"odinger equation~\eqref{second_order} is the starting point, it suffices to demand that ${\cal A}^\dagger\,{\cal A}$ be Hermitian, which leads to the condition that, for any pair of wavefunctions $\psi$ and $\chi$,
\beq{}{}{}{}{}
\psi^{*} \,\partial_z\,\chi \ - \ \partial_z\,\psi^{*} \,\chi \label{bc_second_order}
\eeq
vanish at the boundary. As a result~\eqref{bc_chi} is a possible choice, but one can also impose the boundary condition
\beq{}{}{}{}{}
\cos\theta\,\chi \ + \ \sin\theta\,\partial_z\,\chi \ = \ 0
\eeq
at the boundary. This family of choices contains, as special cases, the familiar Dirichlet and Neumann ones, which are selected by the no-flow conditions of~\cite{boundary}.

We have often resorted to a positivity argument for the operator ${\cal A}^\dagger \,{\cal A}$. This requires that
\beq{}{}{}{}{}{}
\left(\chi, {\cal A}^\dagger\,{\cal A}\,\chi\right) \ = \ \left({\cal A}\,\chi,{\cal A}\,\chi\right) \ ,
\eeq
and this condition demands that, at the boundaries,
\beq{}{}{}{}{}{}
\chi\,{\cal A}\,\chi \ = \ 0 \ .
\eeq
\end{appendices}

\end{document}